\documentclass[preprint,1p,12pt]{elsarticle}

\makeatletter
\def\ps@pprintTitle{
  \let\@oddhead\@empty
  \let\@evenhead\@empty
  \let\@evenfoot\@oddfoot
}
\makeatother

\usepackage{nicefrac}
\usepackage{amssymb,amsmath}
\usepackage[latin1]{inputenc}
\usepackage{color}

\newcommand{\ot}{\leftarrow}
\newcommand{\adjoint}{\mathop{\&}\nolimits}

\renewcommand{\P}[1]{\mathbb{P} #1}

\renewcommand{\Dot}{\hbox{\boldmath $.$}}
\newcommand{\Dotted}[1]{\mathop{\stackrel\Dot{#1}}\nolimits}

\def\P{\ensuremath{\mathbb{P}}}

\journal{Fuzzy Sets and Systems}

\newtheorem{theorem}{Theorem}
\newtheorem{corollary}[theorem]{Corollary}

\newtheorem{proposition}[theorem]{Proposition}
\newdefinition{definition}[theorem]{Definition }
\newdefinition{remark}[theorem]{Remark }
\newdefinition{example}[theorem]{Example }
\newproof{proof}{Proof}

\begin{document}
\begin{frontmatter}

\title{Extended multi-adjoint  logic programming}

\author
{M. Eugenia Cornejo, David Lobo, Jes\'{u}s Medina}

\address
{Department of Mathematics,
 University of  C\'adiz. Spain\\
\texttt{\{mariaeugenia.cornejo,david.lobo,jesus.medina\}@uca.es}}

\begin{abstract}
Extended multi-adjoint logic programming arises as an extension of multi-adjoint normal logic programming where constraints   and a special type of aggregator operator have been included. The use of this general aggregator operator permits to consider, for example, different negation operators in the body of the rules of a logic program.  We have introduced the syntax and the semantics of this new paradigm, as well as an interesting mechanism for obtaining a multi-adjoint normal logic program from an extended multi-adjoint logic program. This mechanism will allow us to establish technical properties relating the different stable models of both logic programming frameworks. Moreover, it  makes possible that the already developed and future theory associated with stable models of multi-adjoint normal logic programs can be applied to extended multi-adjoint logic programs.

\end{abstract}


\end{frontmatter}

\section{Introduction}\label{intro}
Multi-adjoint logic programming is an interesting logical theory introduced by Medina et al.~\cite{lpnmr01} which has gained a lot of popularity
~\cite{Bustince_2015,fss:manlp2017,citagines3,JulianIranzo201727,Madrid_2013,Morcillo2018,Moreno2019,susana-eusflat09}. This logical theory arises as a generalization of different non-classical logic programming approaches~\cite{DP01:ecsqaru,vojtas-fss}, removing their particular details and preserving only the minimal mathematical requirements in order to guarantee the operability. Specifically, a multi-adjoint logic program is defined from an algebraic structure composed of a lattice together with different conjunctors and implications making up adjoint pairs. Notice that the conjunctors do not need to be neither commutative nor associative. 

The multi-adjoint logic programming framework has been recently extended by considering a negation operator in the logic programs, which has given rise to a new logic programming approach called multi-adjoint normal logic programming~\cite{Cornejo:ipmu2018,fss:manlp2017,Madrid_2013}. One of the most interesting properties of multi-adjoint normal logic programs is associated with the existence of stable models, which allows us to check whether the logic program is related to a solvable problem. Note that, when multi-adjoint normal logic programs correspond to some search problem, the stable models are actually its possible solutions. A detailed study on the syntax and semantics of multi-adjoint normal logic programs, including important results about the existence and the unicity of stable models, was introduced in~\cite{fss:manlp2017}. Now, we are interested in  extending the multi-adjoint logic normal programming framework in order to increase its  flexibility and the range of real applications.  

Concerning the syntax of multi-adjoint normal logic programs, the main novelty presented in this paper will be the inclusion of different negation operators in the body of the rules and the consideration of a special type of rule called constraint. This new and more versatile  logic  framework  will be called extended multi-adjoint logic programming and it can be considered as one further step in order to allow an easier transformation from the natural language to decision rules. 

The classical notion of a constraint is related to a rule whose head is empty (or only contains the bottom element) and it is used to specify that the body of the rule should not be satisfied for any valid solution. This definition was extended to the fuzzy case in order to consider in the head any element in the lattice~\cite{Janssen2012}. This consideration enriches the language of the multi-adjoint framework and so, this special kind of rules will be adapted to this framework. On the other hand, taking into account the multi-adjoint philosophy, one can expect that different negation operators appear in the body of the rules of a logic program. This fact will be simulated by the use of general operators which are order-preserving in some arguments and order-reversing in the remainders. 
 
In what regards to the semantics of extended multi-adjoint logic programs, we will continue with the stable model semantics. Based on the motivation and work developed by Janssen et al.~\cite{Janssen2012}, we have introduced a mechanism to translate an arbitrary extended multi-adjoint logic program into an extended multi-adjoint logic program without constraints, which does not increase the number of rules of the original program. Additionally, a procedure to translate an arbitrary extended multi-adjoint logic program without constraints into a multi-adjoint normal logic program will be also given. Both mechanisms will allow us to establish the relationships among the stable models of each mentioned logic programs. From these relationships, we will be capable of ensuring that  there exists a one-to-one correspondence between the stable models of an extended multi-adjoint logic program and  the multi-adjoint normal logic program obtained from the translation mechanisms.

In summary, the advances provided by this work have a strong impact on the syntax and the semantics of the multi-adjoint logic programming framework. The possibility of including different negation operators constraints in the body of the rules and considering constraints makes the rules more flexible and therefore, it makes easier the process of translating the information contained in a text or in a database into a logical program. The more flexible the rules, the easier it will be to translate the information into decision rules and the easier interpretation of those decision rules will be. It is also essential to highlight the importance of the theoretical results developed in this work. This fact lies in the possibility of translating different results given in the multi-adjoint framework~\cite{fss:manlp2017}  or in other more particulars environments, as the residuated one~\cite{Madrid2017,Madrid:2012}, into the new flexible framework of extended multi-adjoint logic programs.

This paper is laid out in the following way. Section~\ref{sec:MANLP} recalls the main concepts and results related to the multi-adjoint normal logic programming framework. Section~\ref{sec:EMALP} presents the syntax and semantics of extended multi-adjoint logic programs. The procedure for translating an arbitrary extended multi-adjoint logic program into an extended multi-adjoint logic program without constraints is included in Section~\ref{sec:noct}, whereas the one given for translating an arbitrary extended multi-adjoint logic program without constraints into a multi-adjoint normal logic program is presented in Section~\ref{sec:final}. The technical properties introduced in Sections~\ref{sec:noct} and~\ref{sec:final} allow us to obtain a multi-adjoint normal logic program with the same stable models as a given extended multi-adjoint logic program. Section~\ref{sec:conclusion} ends with some conclusions and prospects for future work.

\section{Multi-adjoint normal logic programming}\label{sec:MANLP}

Multi-adjoint normal logic programming was introduced in~\cite{fss:manlp2017} as a non-monotonic multi-adjoint logic programming framework in which the considered algebraic structure is enriched with a negation operator. The syntax and the semantics of multi-adjoint normal logic programs are recalled in the sequel.

First of all, we will include the notion of adjoint pair, which arises as a generalization of a t-norm and its residuated implication.
\begin{definition} \label{def:adjointpair}
Let $( P,\leq) $ be a partially ordered set and $(\adjoint{},\ot)$   a pair of binary operations in $P$, we say that  $(\adjoint{},\ot)$ forms an \emph{adjoint pair} in $(P,\leq ) $, if the following properties hold:
\begin{enumerate}
\item $\adjoint$ is order-preserving\footnote{Order-preserving, monotonic and increasing mappings are equivalent notions.} in both arguments, that is, \ if
$x_1,x_2,x, y_1, y_2, y\in P$ and $x_1 \leq  x_2$, $y_1 \leq  y_2$, then $(x_1\adjoint{} y ) \leq  (x_2 \adjoint{} y)$ and $(x\adjoint y_1)\leq  (x \adjoint y_2)$.
\item $\ot$ is order-preserving in the first  argument (the consequent) and order-reserving in the second argument (the antecedent). That is, if $y_1,y_2, y, z_1,z_2, z \in P$  and $y_1 \leq  y_2$, $z_1 \leq z_2$,  then $(z_1 \ot y)
\leq (z_2 \ot y)$,  $(z\ot y_2 )\leq (z \ot y_1)$.

\item $(\adjoint{},\ot)$  satisfies the adjoint property, that is, 
$$
x \leq  (z \ot y) \quad \hbox{  if and only if }\quad (x \adjoint{} y) \leq  z
$$
for all $x,y,z \in P$.
\end{enumerate}
\end{definition}

Example of adjoint pairs defined on $[0,1]$ are the G\"odel, product and \L ukasiewicz t-norms together with their residuated  implications:
  $$
  \begin{array}{lll}
x \adjoint_{\text{G}} y=\min\{x,y\}  &  & z\ot_{\text{G}}y =  \begin{cases}1 & \hbox{if }  y\le z\\z&\hbox{otherwise}\end{cases} \\[3ex]
x \adjoint_{\text{P}} y= x \cdot y &  &   z\ot_{\text{P}}y = \min\{1,z/y\} \\[2ex]
x \adjoint_{\text{\L}} y=\max\{0,x+y-1\} & & z\ot_{\text{\L}}y =\min\{1,1-y+z\}  \\[2ex]
  \end{array}
  $$

The algebraic structure on which multi-adjoint normal logic programs are defined is usually known as multi-adjoint normal lattice. This notion is formally stated as follows.

\begin{definition}\label{multi-adjoint}
The tuple $(L,\preceq,\leftarrow_1,\adjoint_1,\dots,\leftarrow_n,\adjoint_n,\neg)$ is a \emph{multi-adjoint normal lattice}, if the following properties are verified:
\begin{enumerate}
\item $( L,\preceq ) $ is a bounded lattice, i.e.\ it has a bottom $(\bot)$ and a top $(\top)$ element;
\item $(\adjoint_i,\leftarrow_i)$ is an adjoint pair in $( L,\preceq)$, for $i\in\{1,\dots,n\}$;
\item $\top \adjoint_i \vartheta = \vartheta \adjoint_i \top=\vartheta$, for all $\vartheta\in L$ and $i\in\{1, \dots, n\}$;
\item $\neg\colon L\rightarrow L$ is a negation operator, that is, an order-reversing mapping satisfying the equalities $\neg(\bot)=\top$ and \,$\neg(\top)=\bot$.
\end{enumerate}
\end{definition}

Once the underlying structure has been established, a multi-adjoint normal logic program is defined as a set of rules where different implications may appear in different rules of the program, and the body of the rules is composed of aggregation  operators, which will usually be denoted as $@$,  together with a negation operator. In this paper, the usual boundary conditions of the aggregator operator will be not required, hence, the monotonicity will be the needed basic property.
Notice that particular cases of aggregators are conjunctive operators (which in particular can be the adjoint conjunctors of the implications,  that is, $\&_1,  \&_2,  \ldots,
\&_n $), disjunctive operators (denoted as $\vee$), average and hybrid operators, etc. Moreover, it can be the composition of different of these operators. In this case, we will write  the propositional symbols $q_1, \dots, q_n$ appearing in the body of the rule   between brackets, that is, $@[ q_1, \dots, q_n]$. The set of propositional symbols will be denoted as $\Pi$. Therefore, for example, the expression $(p\adjoint_\text{P} q)\vee_{\text{\L}} r$, depending on  $p,q,r\in \Pi$, can be written from the operator ${@}\colon \Pi\times \Pi \rightarrow \Pi$, defined as $@[p,q,r]=(p\adjoint_\text{P} q)\vee_{\text{\L}} r$, for all $p,q,r\in \Pi$. For more details see~\cite{lpnmr01}.

\begin{definition}\label{def:nmalp}
Let $(L,\preceq,\leftarrow_1,\adjoint_1,\dots,\leftarrow_n,\adjoint_n,\neg )$ be a multi-adjoint normal lattice.  A \emph{multi-adjoint normal logic program (MANLP)} $\P$ is a finite set of weighted rules of the form:
$$\langle p\leftarrow_i @[ p_1, \dots, p_m,\neg p_{m+1},\dots,\neg p_n]; \vartheta \rangle$$
where $i\in\{1,\dots,n\}$, $@$ is an aggregator operator, $\vartheta$ is an element of $L$ and $p,p_1,\dots,p_n$ are propositional symbols such that $p_j\neq p_k$, for all $j,k\in\{1,\dots,n\}$, with $j\neq k$. The propositional symbol $p$ is called \emph{head} of the rule, $@[ p_1, \dots, p_m,\neg p_{m+1},\dots,\neg p_n]$ is called \emph{body} of the rule and the value $\vartheta$ is its \emph{weight}.
\end{definition}

The set of propositional symbols appearing in a MANLP $\P$ is usually denoted as $\Pi_\P$. As far as the semantics of MANLPs is concerned, it is based on the stable model semantics. In the following, the notion of interpretation is presented, which plays a fundamental role in the stable models semantics.

\begin{definition}
Given a complete lattice $( L,\preceq ) $, a mapping $I\colon \Pi_\P\to L$, which assigns to every propositional symbol appearing in $\Pi_\P$  an element of the lattice $L$, is called \emph{$L$-interpretation}. The set of all $L$-interpretations is denoted as $\cal{I}_\mathfrak{L}$.
\end{definition}

In order to distinguish a syntactical symbol in a rule from the operator that it represents, we will fix the next notation. Given a symbol $\omega$, its interpretation under a multi-adjoint normal lattice will be denoted as $\Dotted \omega$. Likewise, the evaluation of a formula $\cal B$ under an interpretation $I$ will be given by the uniquely extended interpretation    $\hat{I}({\cal B})$ defined from $I$ using the unique homomorphic extension theorem. 

Considering this notation, the notions of model and satisfiability are defined below.

\begin{definition}\label{def:model}
Given  a MANLP  $\mathbb{P}$ and an interpretation $I \in \cal{I}_\mathfrak{L}$\@, we say that:
\begin{itemize}
\item[(1)] A weighted  rule $\langle p\leftarrow_i @[ p_1, \dots, p_m,\neg p_{m+1},\dots,\neg p_n]; \vartheta \rangle$ in $\P$
is \emph{satisfied} by $I$ if and only if 
\[\vartheta\preceq \hat{I}\left(  p\leftarrow_i @[ p_1, \dots, p_m,\neg p_{m+1},\dots,\neg p_n]\right)\]
\item[(2)]An $L$-interpretation $I \in \cal{I}_\mathfrak{L}$ is a \emph{model} of   $\mathbb{P}$ if and only if all
weighted rules in $\mathbb{P}$ are  satisfied by~$I$\@.
\item[(3)]An $L$-interpretation $I \in \cal{I}_\mathfrak{L}$ is the \emph{least model} of $\mathbb{P}$, if the inequality $I(p) \preceq J(p)$ holds, for all  $p \in \Pi_\P$ and for each model  $J \in\cal{I}_\mathfrak{L}$ of $\mathbb{P}$.
\end{itemize}
\end{definition}

The following definition recalls  the immediate consequence operator for the multi-adjoint normal logic programming framework considered in~\cite{fss:manlp2017}.

\begin{definition}
Let $\P$ be a MANLP. The \emph{immediate consequence operator} is the mapping $T^\mathfrak{L}_\mathbb{P}\colon{\cal I}_\mathfrak{L}\rightarrow {\cal I}_\mathfrak{L}$ defined for every $L$-interpretation $I$ and $p\in\Pi_\P$ as follows:
$$
T_\P(I)(p)=\sup\{\vartheta\Dotted{\adjoint_i} \hat{I}(\mathcal{B}) \mid \langle p\leftarrow_i \mathcal{B};\vartheta \rangle\in\P \}
$$
\end{definition}

It is convenient to mention that, when $\P$ does not contain any negation operator~\cite{lpnmr01},  we obtain that $T_\P$ is monotonic and it has a least fix-point by Knaster-Tarski fix-point theorem~\cite{tarski1955}. From these facts, we can deduce that this least fix-point is the least model of $\P$. For more details, see~\cite{lpnmr01}.

However, the immediate consequence operator is not necessarily monotonic in MANLPs and consequently, the existence of the least model cannot be ensured. For that reason, the semantics of MANLPs is based on a special type of models called stable models. This notion is closely related to the reduct of a MANLP with respect to a given interpretation. Namely, given a MANLP $\P$ and an $L$-interpretation $I$, the reduct of $\P$ with respect to $I$, denoted as $\P_I$, is defined by substituting each rule 
\[\langle p\leftarrow_i @[ p_1, \dots, p_m,\neg p_{m+1},\dots,\neg p_n]; \vartheta \rangle\]
in $\P$ by the rule 
\begin{equation}\label{eq:reductMANLP}
\langle p\leftarrow_i @_{I}[ p_1, \dots, p_m]; \vartheta \rangle
\end{equation}
where the operator $\Dotted@_{I}\colon L^m \rightarrow L$ is defined as 
\[\Dotted@_{I}[ \vartheta_1, \dots, \vartheta_m]\!=\!\Dotted @[\vartheta_1,\dots,\vartheta_m,\Dotted \neg {I}(p_{m+1}),\dots, \Dotted \neg{I}(p_n)]\]
for all $\vartheta_1,\dots,\vartheta_m\in L$.

The concept of stable model is then defined as follows.

\begin{definition}\label{def:stable}
Given  a MANLP $\P$ and an  $L$-interpretation $I$, we say that $I$ is a \emph{stable model} of $\P$ if and only if $I$ is the least model  of $\P_I$.
\end{definition}

It is important to highlight that the semantics of MANLPs is defined in terms of the stable models of the program. Therefore, ensuring the existence of stable models becomes a crucial task in order to define the semantics of a MANLP. The next result provides a sufficient condition to come to this target. In particular, it is related to the continuity of the operators involved in the rules of the MANLP.

\begin{theorem}[\cite{fss:manlp2017}] \label{thm:existencia}
Let $(K,\preceq,\leftarrow_1,\adjoint_1,\dots,\leftarrow_n,\adjoint_n,\neg )$ be a multi-adjoint normal lattice, where $K$ is a non-empty convex compact subset of an euclidean space, and $\P$ be a finite MANLP defined on this lattice. If $\adjoint_1,\dots, \adjoint_n$, $\neg$ and the aggregator operators in the body of the rules of  $\P$ are continuous operators, then $\P$  has at least a stable model.
\end{theorem}

After presenting the main notions corresponding to the multi-adjoint normal logic programming framework, we will focus on extending this logic programming environment to a more general one, including several negation operators in the body of the rules and a new type of rules called constraints. From now on, we will consider  a fixed MANLP $\P$ defined on a multi-adjoint normal lattice $(L,\preceq,\leftarrow_1,\adjoint_1,\dots,\leftarrow_n,\adjoint_n,\neg )$.

\section{Extending multi-adjoint normal logic programs}\label{sec:EMALP}

It is very usual in real cases that some property, attribute or characteristic be limited by an upper bound. Indeed, this boundary value can be given by a combination of   properties. This situation can be simulated in logic programming by the use of constraints. Specifically, in fuzzy logic programming,   
given a program $\P$, a formula $\mathcal{B}$ and an upper bound $c\in L$, we need to ensure that  the inequality $M(\mathcal{B})\preceq c$ holds for every stable model $M$ of $\P$.

In order to capture this condition  in multi-adjoint logic programming,  we propose the inclusion of the rule $\langle c\leftarrow_i \mathcal{B};\ \top\rangle$ in the program $\P$, being $(\adjoint_i,\leftarrow_i)$ any adjoint pair belonging to the multi-adjoint normal lattice in which the program $\P$ is defined, that is, $i\in\{1,\dots,n\}$. This kind of rules, in which an element of the carrier of the multi-adjoint normal lattice appears in the head of the rule, are usually known as \emph{constraints}.   

Next, we will see that the inclusion of constraints provides the required upper bound limitation.
If a stable model $M$ of the program $\P$ satisfies the rule $\langle c\leftarrow_i \mathcal{B};\ \top\rangle$, then the inequality  $\top\preceq M(c\leftarrow_i \mathcal{B})$ should be satisfied, which is equivalent to   $M(\mathcal{B})\Dotted\adjoint_i\top\preceq c$, by the adjoint property.  Since $\top$ is the identity element of $\adjoint_i$, we obtain the inequality\footnote{Notice that we consider $M(c)=c$, for each $c\in L$, as it is formally stated further in Definition~\ref{def:interpretation}.} $M(\mathcal{B})\preceq c$. Therefore, the satisfiability  of the constraint $\langle c\leftarrow_i \mathcal{B};\ \top\rangle$ provides that any  stable model $M$ of $\P$ verifies the inequality $M(\mathcal{B})\preceq c$.

On the other hand, according to the multi-adjoint philosophy, the inclusion of different negation operators in the body of the rules would also be interesting.
This proposal will be carried out by using a special type of aggregator operator in the body of the rules being order-preserving in some arguments and order-reversing in the remainders. 

Hence, hereinafter, an extended multi-adjoint logic programming framework which includes both constraints and aggregator operators with order-reversing arguments will be developed. Namely, this kind of aggregator operators will be called extended aggregators.

\begin{definition}
Let $(L,\preceq)$ be a complete lattice. An \emph{extended aggregator} $@^e\colon L^n\to L$ is any order-preserving mapping on the first $i$-th arguments when $i\in\{1,\dots,m\}$ and order-reversing on the $j$-th arguments when $j\in\{m+1,\dots,n\}$. The image of an element $(x_1,\dots,x_n)\in L^n$ under $@^e$ will be denoted as 
\[@^e(x_1,\dots,x_n)=@^e[x_1,\dots,x_m;x_{m+1},\dots,x_n]\]

A multi-adjoint lattice $(L,\preceq,\leftarrow_1,\adjoint_1,\dots,\leftarrow_n,\adjoint_n)$ enriched with a family of extended aggregators $@^e_1,\dots, @^e_k$, will be called \emph{extended multi-adjoint lattice}. 
\end{definition}

This general algebraic structure together with constraints are the pillars from which we will extend MANLP to a more flexible framework. 

\begin{definition}\label{def:EMALP}
Let $(L,\preceq,\leftarrow_1,\adjoint_1,\dots,\leftarrow_n,\adjoint_n,@^e_1,\dots, @^e_k)$ be an extended multi-adjoint lattice.  An \emph{extended multi-adjoint logic program (EMALP)} $\P^e$ is a finite set of weighted rules of the form
\[\langle p\leftarrow_i @^e[ p_1, \dots, p_m; p_{m+1},\dots, p_n]; \vartheta \rangle\]
and rules of the form
\[\langle c\leftarrow_i @^e[ p_1, \dots, p_m; p_{m+1},\dots, p_n]; \top \rangle\]
where $i\in\{1,\dots,n\}$, $@^e\in\{@^e_1,\dots,@^e_k\}$, $\vartheta$ and $c$ are elements of $L$ and $p,p_1,\dots,p_n$ are propositional symbols such that $p_{s_1}\neq p_{s_2}$, for all ${s_1},{s_2}\in\{1,\dots,n\}$, with ${s_1}\neq {s_2}$.
\end{definition}

Notice that, the extended aggregators considered in the extended multi-adjoint lattice and   used in the program,  can be obtained from the composition of different monotonic operators (such as, conjunctors and disjunctions) and negations operators.

As it was mentioned at the beginning of this section, we can simulate more than just one negation operator. Namely, consider a multi-adjoint lattice $(L,\preceq,\leftarrow_1,\adjoint_1,\dots,\leftarrow_n,\adjoint_n)$ enriched with the negation operators $\neg_1,\dots,\neg_k$, and a set of rules of the form
\[\langle p\leftarrow_i @[ p_1, \dots, p_m,\neg_{j_{m+1}} p_{m+1},\dots,\neg_{j_n} p_n]; \vartheta \rangle\]
where $i\in\{1,\dots,n\}$, $j_l\in\{1,\dots,k\}$ for each $l\in\{m+1,\dots,n\}$, $@$ is an aggregator, $\vartheta\in L$ and $p,p_1,\dots,p_n$ are propositional symbols such that $p_{s_1}\neq p_{s_2}$, for all ${s_1},{s_2}\in\{1,\dots,n\}$, with ${s_1}\neq {s_2}$.

Then, the MANLP $\P$ composed of such set of rules can be rewritten as an EMALP as follows. For each rule $r$ of the form
\[\langle p\leftarrow_i @[ p_1, \dots, p_m,\neg_{j_{m+1}} p_{m+1},\dots,\neg_{j_n} p_n]; \vartheta \rangle\]
with $j_l\in\{1,\dots,k\}$ for each $l\in\{m+1,\dots,n\}$, occurring in $\P$, we consider the rule $r^e$ given by
\[\langle p\leftarrow_i @^e[ p_1, \dots, p_m; p_{m+1},\dots, p_n]; \vartheta \rangle\]
being
\[@^e[ p_1, \dots, p_m; p_{m+1},\dots, p_n]=@[ p_1, \dots, p_m,\neg_{j_{m+1}} p_{m+1},\dots,\neg_{j_n} p_n]\]
Due to the fact that $@$ is an aggregator and $\neg_1,\dots,\neg_k$ are negation operators, the mapping $@^e$ is undeniably an extended aggregator. Hence, the program
\[\P^e=\{r^e\mid r\in\P\}\] 
is clearly an EMALP. Therefore, if we want to consider MANLPs in which more than one negation operator appears, we can do it by means of EMALPs.

In order to illustrate the expressivity power of the extended multi-adjoint logic programming framework, a practical toy example will be developed in the sequel.

\begin{example}
A group of experts stated that the level of water, the level of oil and the temperature are critical features for the acceptable behaviour of a motor. Specifically, they reached the following conclusions:
\begin{itemize}
\item[(a)] The level of water and the level of oil seriously affect to the temperature of the motor. In particular, if either the level of water or the level of oil is low, then the temperature of the motor increases significantly, being the level of water more damaging than the level of oil.
\item[(b)] The level of oil and the temperature of the motor must be controlled, since in case that these two features are high at the same time, then the motor is broken.
\end{itemize}
Consider the variables $w,o,t\in[0,1]$, which represent the level of water, the level of oil and the temperature of the motor, respectively. Namely, the value $1$ (respectively $0$) for the variable w/o/t corresponds to a high (respectively low) level of water/level of oil/temperature.

The behaviour detailed in Statement (a) can be modelled by means of the following rule:
\[\langle t\leftarrow_\text{P} \max\{ \neg_1(o),\neg_2(w)\}; 0.9 \rangle\]
where the negation operators $\neg_1,\neg_2\colon[0,1]\to[0,1]$ are defined as $\neg_1(x)=1-x$ and $\neg_2(x)=(1-x^2)^{1/2}$, for all $x\in[0,1]$.

Observe that, the rule has a high weight due to the level of water and the level of oil have a big impact on the temperature of the motor. We make use of different negations in order to model that this impact is different, being stronger in the case of low water than in the case of low oil. Indeed, $\neg_1(x)\leq\neg_2(x)$, for all $x\in[0,1]$. Finally, 
we consider the maximum between  $\neg_1(o)$ and $\neg_2(w)$, because a low level of any of these two variables, oil and water,  is enough in order to increase the temperature of the motor.

In what regards Statement (b),  since the motor cannot work with a high level of oil and a high temperature, then we need to include this requirement through the following constraint:
\[\langle 0.8\leftarrow_\text{P} o\adjoint_\text{P} t; 1 \rangle\]
in which  the threshold $0.8$ prevents the level of oil and the temperature of the motor, from being high at the same time. In this case, the operator $\adjoint_\text{P}$ has ben selected in order to represent the conjunction of the natural language. 
Notice that, the implication can be any residuated implications because the truth value  is the top element $1$. 
\qed
\end{example}

As far as the semantics of EMALPs is concerned, the notions of interpretation, satisfaction of a rule and model are defined in a similar way to the semantics of MANLPs.

\begin{definition}\label{def:interpretation}
Let $\P^e$ be an EMALP on an extended multi-adjoint lattice $(L,\preceq,\leftarrow_1,\adjoint_1,\dots,\leftarrow_n,\adjoint_n,@^e_1,\dots, @^e_k)$. An interpretation of $\P^e$ is any mapping $I^e\colon \Pi_{\P^e}\to L$, where $\Pi_{\P^e}$ is the set of propositional symbols appearing in $\P^e$. The set of interpretations of $\P^e$ is usually denoted as $\mathcal{I}_{\mathfrak{L}}^e$.
\end{definition}

As usual, the evaluation of a formula under an interpretation proceeds inductively. For the sake of completeness, an interpretation $I^e$ is extended to the set $L$ as $I^e(x)=x$ for each $x\in L$.

Basing on the complete lattice $(L,\preceq)$, an order relation can be defined in the set of interpretations of $\P^e$. Namely, given two interpretations $I^e_1, I^e_2\in\mathcal{I}_{\mathfrak{L}}^e$, we say that $I^e_1 \sqsubseteq I^e_2$ if and only if $I^e_1(p) \preceq I^e_2(p)$, for all $p \in \Pi_{\P^e}$.

\begin{definition} Given  an EMALP $\P^e$  and an interpretation $I^e \in \mathcal{I}_{\mathfrak{L}}^e$\@, we say that:
\begin{itemize}
\item[(1)] A rule $\langle p\leftarrow_i @^e[ p_1, \dots, p_m; p_{m+1},\dots, p_n]; \vartheta \rangle$
of $\P^e$ is \emph{satisfied} by $I^e$ if and only if $\vartheta\preceq \hat{I^e}\left( p\leftarrow_i @^e[ p_1, \dots, p_m; p_{m+1},\dots, p_n]\right)$.

Likewise, a constraint $\langle c\leftarrow_i @^e[ p_1, \dots, p_m; p_{m+1},\dots, p_n]; \vartheta \rangle$ of $\P^e$ is \emph{satisfied} by $I^e$ if and only if $\vartheta\preceq c\Dotted\leftarrow_i \hat{I^e}\left(@^e[ p_1, \dots, p_m; p_{m+1},\dots, p_n]\right)$.
\item[(2)]An interpretation $I^e \in \mathcal{I}_\mathfrak{L}^e$ is a \emph{model} of $\P^e$ if and only if all
weighted rules in $\mathbb{P}$ are  satisfied by~$I^e$\@.
\end{itemize}
\end{definition}

Similarly to the multi-adjoint normal logic programming framework, the notion of stable model for EMALPs is stated by means of the concept of reduct with respect to an interpretation. In this case, given an interpretation, each rule in the EMALP is substituted in the reduct by an analogous rule in which the atoms occurring in any order-reversing argument of the aggregator in the body of the rule are evaluated under the given interpretation.

Formally, given an EMALP $\P^e$ and an interpretation $M^e$, the reduct of $\P^e$ with respect to $M^e$, denoted as $\P^e_{M^e}$, is built substituting each rule in $\P^e$
\[\langle l\leftarrow_i @^e[ p_1, \dots, p_m; p_{m+1},\dots, p_n]; \vartheta \rangle\] 
 where $l$ is either a propositional symbol or a value in $L$, and $@^e\in\{@^e_1,\dots, @^e_k\}$, by the rule
\[\langle l \leftarrow_i @^e_{M^e}[ p_1, \dots, p_m]; \vartheta \rangle\]
where the aggregator $\Dotted {@^e}_{M^e}\colon L^m\to L$ is defined as 
\begin{equation}\label{eq:reduct}
\Dotted {@^e}_{M^e}[\vartheta_1,\dots,\vartheta_m]=\Dotted {@^e}[\vartheta_1,\dots,\vartheta_m;M^e(p_{m+1}),\dots,M^e(p_n)]
\end{equation}
Evidently, as $@^e$ is an order-preserving mapping in the first $m$ arguments,  $@^e_{M^e}$ is an aggregator operator.
Hence, we conclude that the program $\P^e_{M^e}$ is a monotonic multi-adjoint logic program. The notion of stable model is then established as follows

\begin{definition}
Let $\P^e$ be an EMALP. An interpretation $M^e$ is said to be a stable model of $\P^e$ if $M^e$ is the least model of $\P^e_{M^e}$.
\end{definition}

The following proposition shows an interesting property of stable models.

\begin{proposition}\label{prop:2.11NicoyManolo}
Any stable model of an EMANLP $\P^e$ is a minimal model of~$\P^e$.
\end{proposition}
\begin{proof}
	Consider that $M^e$ is a stable model of $\P^e$. In other words, $M^e$ is the least model of the program $\P^e_{M^e}$. We will prove by reductio ad absurdum that $M^e$ is a minimal model of $\P^e$.
	
	Thus, suppose that there exists a model $N^e$ of $\P^e$ such that $N^e\sqsubset M^e$. We will see that $N^e$ is a model of $\P^e_{M^e}$.
	
	Since $N^e$ is a model of $\P^e$, for each rule in $\P^e$ of the form
	\[\langle l\leftarrow_i @^e[ p_1, \dots, p_m; p_{m+1},\dots, p_n]; \vartheta \rangle\]
	 where $l$ is either a propositional symbol or a value in $L$, we obtain that
	\[\vartheta\preceq \hat{N^e}\left( l\leftarrow_i @^e[ p_1, \dots, p_m; p_{m+1},\dots, p_n]\right)\]
	Equivalently,
	\[\vartheta \preceq N^e(l)\Dotted\leftarrow_i \Dotted {@^e}[N^e(p_1),\dots,N^e(p_m); N^e(p_{m+1}),\dots, N^e(p_n)]\]
	Now, as $\Dotted {@^e}$ is order-reversing in its last $n-m$ arguments and $N^e\sqsubset M^e$, the following inequality holds:
	\[{\footnotesize \Dotted {@^e}[N^e(p_1),\dots,N^e(p_m); M^e(p_{m+1}\!),\dots, M^e(p_n)]\preceq\Dotted {@^e}[N^e(p_1),\dots,N^e(p_m); N^e(p_{m+1}\!),\dots, N^e(p_n)]}\]
	Therefore, according to the fact that $\leftarrow_i$ is order-reversing in the antecedent, we deduce that
	\begin{eqnarray*}
		\vartheta&\preceq&N^e(l)\Dotted\leftarrow_i \Dotted {@^e}[N^e(p_1),\dots,N^e(p_m);  N^e(p_{m+1}),\dots,  N^e(p_n)]\\
		&\preceq&N^e(l)\Dotted\leftarrow_i \Dotted {@^e}[N^e(p_1),\dots,N^e(p_m);  M^e(p_{m+1}),\dots,  M^e(p_n)]\\
		&=&N^e(l)\Dotted\leftarrow_i\Dotted {@^e}_{M^e}[ N^e(p_1),\dots,N^e(p_m)]
	\end{eqnarray*}

As this inequality is true for every rule, we obtain that  $N^e$ is a model of $\P^e_{M^e}$, which leads us to a contradiction, since $M^e$ is the least model of $\P^e_{M^e}$ by hypothesis.
	\qed
\end{proof}

The next example illustrates the concepts defined in this section. In addition, it shed lights on how flexible the extended multi-adjoint logic programming framework is.

\begin{example}\label{ex:EMALP}
Let us consider the extended multi-adjoint lattice given by $([0,1],\leq,\leftarrow_\text{G},\adjoint_G,\leftarrow_\text{P},\adjoint_\text{P},\leftarrow_\text{\L},\adjoint_\text{\L},@^e_1,@^e_2,@^e_3,@^e_4,@^e_5)$, where  the adjoint pairs are the well-known G\"odel, product and \L ukasiewicz pairs~\cite{ija-cmr15} and 
the extended aggregators are defined on $[0,1]^4$  as follows 
\begin{eqnarray*}
@^e_1[x,y;z,t]&=&\min\Big\{\frac{y}{z+t+0.1},1\Big\}\\
@^e_2[x,y;z,t]&=&\max\{\neg_1 z, \neg_2 t\}\\
@^e_3[x;y,z,t]&=&\neg_1 y\\
@^e_4[x,y,z,t]&=&1\\
@^e_5[x,y,z,t]&=&\max\{z,0.7\}
\end{eqnarray*}
for all $x,y,z,t\in [0,1]$, 
and the negation operators $\neg_1,\neg_2\colon[0,1]\to[0,1]$ defined as $\neg_1(x)=1-x$ and $\neg_2(x)=(1-x^2)^{1/2}$, for all $x\in[0,1]$. 

Notice that, the aggregators  $@_1^e$ and $@_2^e$ are both order-preserving on the two first arguments, and order-reversing in the last two arguments. The  aggregator $@_3^e$  is order-reversing on the second argument. Since the rest of variables not appear in the computation, we can consider the semicolon in between $x$ and $y$. We can also consider that the aggregators $@_4^e$ and $@_5^e$ are order-preserving  in all the arguments. Hence, in order to simplify the notation we have written $[x,y,z,t]$ instead of $[x,y,z,t;]$. 

In this setting, we consider the  EMALP  $\P^e$ 
 composed of the following three weighted rules, one constraint and one fact.
 \begin{equation*}
\begin{array}{ll}
r_1^e:\ \langle p\leftarrow_\text{P} @^e_1[p,q;s,t]\ ;\ 0.5\rangle\quad &
r_4^e:\ \langle s\leftarrow_\text{G} @^e_4[p,q,s,t]\ ;\ 0.8\rangle\\
r_2^e:\ \langle q\leftarrow_\text{P}@^e_2[p,q;s,t] \ \ ;\ 0.6\rangle &
r_5^e:\ \langle t\leftarrow_\text{G} @^e_5[p,q,s,t]\ ;\ 0.8\rangle\\
r_3^e:\ \langle 0.7\leftarrow_\text{\L} @^e_3[p;q,s,t]\ ;\ 1\rangle &
\end{array}
\end{equation*}

It is important to highlight the crucial role that the rule $r_3^e$ plays with respect to the value of $q$ in the models of $\P^e$. 
Namely, any interpretation $M^e$ satisfying the rule $r_3^e$ in $\P^e$ verifies the inequality $1\leq \hat{M^e}(0.7\leftarrow_\text{\L}\neg_1 q)$, that is, $1\leq0.7\Dotted\leftarrow_\text{\L} \Dotted\neg_1M^e(q)$. Therefore, according to the adjoint property, satisfied by $(\adjoint_\text{\L},\leftarrow_\text{\L})$, and the definition of $\Dotted\neg_1$, we obtain that $1-M^e(q)\leq0.7$. Equivalently, $0.3\leq M^e(q)$. Hence, the least value that $q$ can take under any model of $\P$ is $0.3$. In other words, we demand the models of $\P$ to satisfy that the evaluation of $q$ is greater or equal  than $0.3$.

For instance, we will see in the following that the interpretation given by $M^e\equiv\{(p,0.25),(q,0.4),(s,0.9),(t,0.85)\}$ is a model of $\P^e$. For the rule $r_1^e$, we obtain that
\begin{eqnarray*}
\hat{M^e}\Big(p\leftarrow_\text{P} \min\Big\{\frac{q}{s+t+0.1},1\Big\}\Big)
&=&M^e(p)\Dotted\leftarrow_\text{P}  \min\Big\{\frac{M^e(q)}{M^e(s)+M^e(t)+0.1},1\Big\}\\
&=&0.25\Dotted\leftarrow_\text{P} \min\Big\{\frac{0.4}{0.9+0.85+0.1},1\Big\}\\
&=&0.25\Dotted\leftarrow_\text{P} \min\Big\{\frac{8}{37},1\Big\}\\
&=&0.25\Dotted\leftarrow_\text{P} \frac{8}{37}=1
\end{eqnarray*}
Therefore $0.5\leq \hat{M^e}\big(p\leftarrow_\text{P} \min\big\{\frac{q}{s+t+0.1},1\big\}\big)$, that is, $M^e$ satisfies the rule $r_1^e$. With regard to the rule $r_2^e$, the following chain of equalities holds
\begin{eqnarray*}
\hat{M^e}(q\leftarrow_\text{P} \max\{\neg_1 s, \neg_2 t\})
&=&M^e(q)\Dotted\leftarrow_\text{P} \max\{\Dotted\neg_1 M^e(s), \Dotted\neg_2 M^e(t)\}\\
&=&0.4\Dotted\leftarrow_\text{P} \max\{\Dotted\neg_1 0.9, \Dotted\neg_2 0.85\}\\
&=&0.4\Dotted\leftarrow_\text{P} \max\{0.1, \frac{\sqrt{111}}{20}\}\\
&=&0.4\Dotted\leftarrow_\text{P}\frac{\sqrt{111}}{20}=\frac{8}{\sqrt{111}}\approx0.76
\end{eqnarray*}
As a consequence, $0.6\leq \hat{M^e}(q\leftarrow_\text{P} \max\{\neg_1 s, \neg_2 t\})$. Thus, the rule $r_2^e$ is satisfied by $M^e$. Moreover, as $0.3\leq0.4=M^e(q)$, the interpretation $M^e$ satisfies the rule $r_3^e$ as well. Similarly, as $0.8\leq0.9=M^e(s)$, $M^e$ satisfies the rule $r^e_4$. Finally, the computations corresponding to the satisfaction of $r_5^e$ are given below.
\begin{eqnarray*}
\hat{M^e}(t\leftarrow_\text{G} \max\{s,0.7\})
&=&M^e(t)\Dotted\leftarrow_\text{G} \max\{M^e(s),0.7\}\\
&=&0.85\Dotted\leftarrow_\text{G} \max\{0.9,0.7\}\\
&=&0.85\Dotted\leftarrow_\text{G} 0.9=0.85
\end{eqnarray*}
Hence, we conclude that the interpretation $M^e$ satisfies all rules appearing in $\P^e$, and thus it is a model of $\P^e$.

In spite of this fact, one can easily check that $M^e$ is not a stable model of the EMALP $\P^e$. Indeed, it is not a minimal model of $\P^e$. For instance, the interpretation $N^e$ given by $N^e\equiv\{(p,\frac{9}{85}),(q,0.36),(s,0.8),(t,0.8)\}$ is also a model of $\P^e$ and clearly satisfies that $N^e\sqsubseteq M^e$. Therefore, applying Proposition~\ref{prop:2.11NicoyManolo}, we deduce that $M^e$ is not a stable model of $\P^e$. In the following, we show that the interpretation $N^e$ is not only a model of $\P^e$, but a stable model. That is, $N^e$ is the least model of the reduct of $\P^e$ with respect to $N^e$, denoted as $\P^e_{N^e}$.

We must recall that the least model of a positive program $\P^e_{N^e}$ is equivalent to the least fix-point of the immediate consequence operator  $T_{\P^e_{N^e}}$~\cite{fss:manlp2017,Ll,lpnmr01}. Therefore, we will prove that  the interpretation $N^e$ is a stable model of $\P^e$,  proving that the least fix-point of $T_{\P^e_{N^e}}$ is $N^e$. 

Following the Knaster-Tarski's theorem, we obtain the least fix-point iterating the 
operator $T_{\P^e_{N^e}}$  from the least interpretation: $I_\bot\equiv\{(p,0),(q,0),(s,0),(t,0)\}$ until a fix-point arises, which is the least one. Hence, from the 
  reduct $\P^e_{N^e}$, which is composed of the rules
 \begin{equation*}
\begin{array}{ll}
r_1^{e^*}:\ \langle p\leftarrow_\text{P} \min\Big\{\frac{y}{s+t+0.1},1\Big\}\ ;\ 0.5\rangle\quad &
r_4^{e^*}:\ \langle s\leftarrow_\text{G} 1\ ;\ 0.8\rangle\\
r_2^{e^*}:\ \langle q\leftarrow_\text{P} \max\{0.2, 0.6\} \ \ ;\ 0.6\rangle &
r_5^{e^*}:\ \langle t\leftarrow_\text{G} \max\{s,0.7\}\ ;\ 0.8\rangle\\
r_3^{e^*}:\ \langle 0.7\leftarrow_\text{\L} 0.64\ ;\ 1\rangle &
\end{array}
\end{equation*}
we compute  the $T_{\P^e_{N^e}}$, obtaining the following iterations:
\begin{center}
\begin{tabular}{|c||c|c|c|c|}
\hline
  & $p$ & $q$ & $s$ & $t$\\[1ex]
\hline
$I_\bot$ & 0 & 0 & 0 & 0\\
\hline
$T_{\P^e_{N^e}}(I_\bot)$ & 0 & 0.36 & 0.8 &0.7\\
\hline
$T_{\P^e_{N^e}}^2(I_\bot)$ & $\nicefrac{9}{80}$ & $0.36$ & 0.8 &0.8\\
\hline
$T_{\P^e_{N^e}}^3(I_\bot)$ & $\nicefrac{9}{85}$ & 0.36 & 0.8 &0.8\\
\hline
$T_{\P^e_{N^e}}^4(I_\bot)$ & $\nicefrac{9}{85}$ & 0.36 & 0.8 &0.8\\
\hline
\end{tabular}
\end{center}
Therefore, after the third iteration  the fix-point arises and we can ensure that $N^e$ is the least model of  $\P^e_{N^e}$ and so, a stable model of $\P^e$.
\qed
\end{example}

The notion of stable model plays a crucial role in the definition of the semantics of an EMALP. Given the set of all stable models of an EMALP, one can check whether a statement is a consequence of the program by simply computing the truth value of the statement on each stable model. Hence, finding conditions which provide information on the number and the form of stable models becomes a critical task in order to define the semantics of an EMALP.

The following sections will focus on obtaining a MANLP equivalent to a given EMALP, from a semantical point of view. We will see that any condition related to the stable models in the multi-adjoint normal logic programming framework can be used in extended multi-adjoint logic programming.

\section{Translating EMALPs  into constraint-free EMALPs.}\label{sec:noct}

The main goal of this section is to obtain a semantically equivalent program to the original one, transforming the constraints to weighted rules. 
Hence, given an EMALP, we will see how to build a new constraint-free EMALP with the same stable models as the given EMALP. 
First of all, we will show that the mechanism given in~\cite{Janssen2012} can be adapted to our framework and then we will improve it avoiding the inclusion of a bigger number of rules. 
 
Given an EMALP $\P^e$ defined on an extended multi-adjoint lattice $(L,\preceq,\leftarrow_1,\adjoint_1,\dots,\leftarrow_n,\adjoint_n, @^e_1,\dots, @^e_k)$, observe that the main obstacle in order to include a constraint  $r$ in $\P^e$  
$$r:\quad\langle c\leftarrow_i @^e[ p_1, \dots, p_m; p_{m+1},\dots, p_n]; \top \rangle$$
in a MANLP is   that $c$ is not a propositional symbol, but an element of the lattice $L$. This obstacle is overcome if there exists a propositional symbol $p_c$ such that any stable model $M^e$ of $\P^e$ satisfies $M^e(p_c)=c$, since in that case the rule $r$ is semantically equivalent to the rule
$$r^*:\quad\langle p_c\leftarrow_i @^e[ p_1, \dots, p_m; p_{m+1},\dots, p_n]; \top \rangle$$
Therefore, the stable models of the EMALP $\P^{\tilde{e}}=\P^e\setminus \{r\}\cup \{r^*\}$ coincide with the stable models of $\P^e$. Hence,  we can include in the program a new propositional symbol $p_c$ such that $M^e(p_c)=c$ for each stable model $M^e$ of $\P^e$. Janssen et al.~\cite{Janssen2012} create such propositional symbol by adding the following two rules in the program:
\begin{eqnarray*}
r_c^1 & : & \quad\langle p_c\leftarrow_j c; \top \rangle\\
r_c^2 & : & \quad\langle p_\bot\leftarrow_j g_\bot(\neg p_\bot)\ \adjoint\ g_c(p_c); \top \rangle
\end{eqnarray*}
where $j\in\{1,\dots,n\}$, $p_\bot$ is a new propositional symbol, $\adjoint$ is a  conjunctor of the multi-adjoint lattice, $\neg$ any negation operator and the mapping $g_c\colon L\to L$ is defined, for each $c\in L$ and $x\in L$, as
\[g_c(x)= \left\{ \begin{array}{lcl}
             \top & \    & \hbox{if}\ c\prec x \\
             \bot & \  & \hbox{otherwise}
             \end{array}
   \right.\]
Notice that $g_c$ is an order-preserving mapping, for each $c\in L$, and thus the mapping $@^e\colon L^2\to L$ given by 
$$@^e[x;y]=g_\bot(\neg y)\adjoint g_c(x)$$
is order-preserving in the first argument and order-reversing in the second argument. Consequently, $@^e$ is an extended aggregator operator and, as a result, $r_c^2$ is a rule which can be included in an extended multi-adjoint logic program. Let $\mathcal{C}_{\P^e}$ be the set of constraints of $\P^e$ and $\mathcal{K}_{\P^e}$ the set of elements of the lattice that occur in the head of constraints, that is $\mathcal{K}_{\P^e}=\{c\mid\langle c\leftarrow_i \mathcal{B};\top\rangle\in\mathcal{C}_{\P^e}\}$, where $\mathcal{B}$ represents the body of a rule. The corresponding constraint-free EMALP $\P^{\tilde{e}}$ of $\P^e$ is then given by
\begin{eqnarray*}
\P^{\tilde{e}}&=&\{r\mid r\in\P^e\setminus\mathcal{C}_{\P^e}\}\\
& \cup & \{\langle p_c\leftarrow_i \mathcal{B};\ \top\rangle\mid\langle c\leftarrow_i \mathcal{B};\ \top\rangle\in\mathcal{C}_{\P^e}\}\\
& \cup & \{\langle p_c\leftarrow_j c; \top \rangle\mid c\in\mathcal{K}_{\P^e}\}\\
& \cup & \{\langle p_\bot\leftarrow_j g_\bot(\neg p_\bot)\ \adjoint\ g_c(p_c); \top \rangle\mid c\in\mathcal{K}_{\P^e}\}
\end{eqnarray*}
with $i,j\in\{1,\dots,n\}$. Following an analogous reasoning to the one given in~\cite{Janssen2012}, it can be proved that the stable models of $\P^e$ are equivalent to the stable models of $\P^{\tilde{e}}$.

Notice that, in order to obtain the constraint-free EMALP $\P^{\tilde{e}}$, the number of new rules that we need to add (for transforming the constraints) to the  EMALP $\P^e$  does not depend on the number of constraints, but in the cardinal of $\mathcal{K}_{\P^e}$, specifically, it  is equal to $|\mathcal{C}_{\P^e}| + 2|\mathcal{K}_{\P^e}|$. As a consequence, the program $\P^{\tilde{e}}$ has $2|\mathcal{K}_{\P^e}|$ more rules than the original EMALP $\P^e$.

In the sequel, we provide a new procedure from which one can build a constraint-free EMALP $\P^{\tilde{e}}$ from $\P^e$ whose number of rules coincides with the number of rules in $\P^e$. Furthermore, this new EMALP only requires one new propositional symbol whilst the strategy suggested in~\cite{Janssen2012} demands $|\mathcal{K}_{\P^e}|+1$ new propositional symbols.

To reach this goal, given $c\in L$, consider the mapping $f_c\colon L\to L$ defined for each $x\in L$ as
\[f_c(x)= \left\{ \begin{array}{lcl}
             \bot & \    & \hbox{if}\ x \preceq c \\
             \top & \  & \hbox{otherwise}
             \end{array}
   \right.\]

Note that, $g_c$ and $f_c$ are different since in a general complete lattice $c\not\prec x$ is not equivalent to $x\preceq c$, since $x$ and $c$ can be incomparables. 

Now, taking into account that  $f_c$ is an order-preserving mapping, for each $c\in L$, we can establish the definition of the corresponding constraint-free EMALP of a given EMALP as follows.

\begin{definition}\label{def:correspEMALP}
Let $\P^e$ be an EMALP and $\mathcal{C}_{\P^e}$ the set of constraints of $\P^e$. The corresponding constraint-free EMALP $\P^{\tilde{e}}$ of $\P^e$ is defined as the following set of rules:
\begin{eqnarray*}
\P^{\tilde{e}}&=&\{r\mid r\in\P^e\setminus\mathcal{C}_{\P^e}\}\\
& \cup & \{\langle p_\bot\leftarrow_i f_\bot(\neg p_\bot)\adjoint f_c(\mathcal{B}); \top\rangle\mid\langle c\leftarrow_i \mathcal{B};\ \top\rangle\in\mathcal{C}_{\P^e}\}
\end{eqnarray*}
where $i\in\{1,\dots,n\}$, $p_\bot$ is a new propositional symbol, $\adjoint$ is any conjunction of the extended multi-adjoint lattice and $\neg$ is any negation operator.
\end{definition}

In the following, we will see that the program built in Definition~\ref{def:correspEMALP} is well-defined, that is, $\P^{\tilde{e}}$ is a   EMALP, which evidently has no   constraint. Notice that, for each rule $r^{\tilde{e}}\in\P^{\tilde{e}}$ of the form
\[r^{\tilde{e}}:\quad\langle p_\bot\leftarrow_i f_\bot(\neg p_\bot)\adjoint f_c(@^e[ p_1, \dots, p_m; p_{m+1},\dots, p_n]);\ \top\rangle\]
as $\neg$ is an order-reversing mapping and, for each $c\in L$, $f_c$ is order-preserving, we deduce that the mapping $\Dotted {@^{\tilde{e}}}\colon L^{n+1}\to L$ defined as
\[\Dotted {@^{\tilde{e}}}[p_1,\dots,p_m;p_{m+1},\dots,p_n,p_\bot]=f_\bot(\Dotted \neg p_\bot)\Dotted\adjoint f_c(\Dotted{@^e}[ p_1, \dots, p_m; p_{m+1},\dots, p_n])\]
is order-preserving in the first $m$ arguments and order-reversing in the last $n-m+1$ arguments. As a result, $@^{\tilde{e}}$ is an extended aggregator operator, and therefore the corresponding program $\P^{\tilde{e}}$ of an EMALP $\P^e$ is, in fact, a constraint-free EMALP.

Now, we will show how the reduct $\P^{\tilde{e}}_{M^{\tilde{e}}}$ of the program $\P^{\tilde{e}}$ with respect to a given interpretation $M^{\tilde{e}}\colon\Pi_{\P^e}\cup\{p_\bot\}\to L$ is defined. Each rule in $\P^{\tilde{e}}$ of the form
\[\langle p_\bot\leftarrow_i f_\bot(\neg p_\bot)\adjoint f_c(@^e[ p_1, \dots, p_m; p_{m+1},\dots, p_n]); \top\rangle\]
is substituted in the reduct $\P^{\tilde{e}}_{M^{\tilde{e}}}$ by the rule
\begin{equation}\label{eq:corr_reduct}
\langle p_\bot\leftarrow_i f_\bot(\hat{M}^{\tilde{e}}(\neg p_\bot))\adjoint f_c(@^e[p_1,\dots,p_m;M^{\tilde{e}}(p_{m+1}),\dots,M^{\tilde{e}}(p_n)]); \top\rangle
\end{equation}
since $\Dotted {@^{\tilde{e}}}$ is order-preserving in the first $m$ arguments and order-reversing in the last $n-m+1$ arguments.

Notice that, each rule in $\P^{\tilde{e}}$ belonging to $\P^e\setminus\mathcal{C}_{\P^e}$ is substituted in the reduct $\P^{\tilde{e}}_{M^{\tilde{e}}}$ following the same procedure to the one given in Section~\ref{sec:EMALP} (Equation~\eqref{eq:reduct}).

The next result shows the existing relation between the stable models of an EMALP and the stable models of its corresponding constraint-free EMALP.

\begin{theorem}\label{th:EMALP}
Let $\P^e$ be an EMALP, $\P^{\tilde{e}}$ the corresponding constraint-free EMALP of $\P^e$,   $M^e\colon\Pi_{\P^e}\to L$ an interpretation, and  $M^{\tilde{e}}\colon \Pi_{\P^e}\cup\{p_\bot\}\to L$ the mapping defined as $M^{\tilde{e}}(p)=M^e(p)$ if $p\in\Pi_{\P^e}$ and $M^{\tilde{e}}(p_\bot)=\bot$. 
We have that 
 $M^e$ is a stable model of $\P^e$ if and only if  $M^{\tilde{e}}$ is a stable model of~$\P^{\tilde{e}}$. 
\end{theorem}
\begin{proof}
Given an interpretation  $N^e\colon\Pi_{\P^e}\to L$, we define $N^{\tilde{e}}\colon \Pi_{\P^e}\cup\{p_\bot\}\to L$ as   $N^{\tilde{e}}(p)=N^e(p)$ if $p\in\Pi_{\P^e}$ and $N^{\tilde{e}}(p_\bot)=\bot$.

First of all, we will see that, given two interpretations $M$ and $N$, we have that  $N^e$ is a model of the reduct $\P^e_{M^e}$ if and only if $N^{\tilde{e}}$ is a model of the reduct $\P^{\tilde{e}}_{M^{\tilde{e}}}$. 

Given a rule $r^e\in\P^e$, we denote by $r^{\tilde{e}}$ its corresponding rule in $\P^{\tilde{e}}$. Observe that, if $r^e\in\P^e\setminus\mathcal{C}_{\P^e}$, then $r^{\tilde{e}}=r^e$. Clearly, as the propositional symbol $p_\bot$ does not occur in the rule $r^e$, it neither does in $r^{\tilde{e}}$, and therefore $N^e$ satisfies the rule $r^e$ if and only if $N^{\tilde{e}}$ satisfies the rule $r^{\tilde{e}}$.

Now, we suppose that $r^e\in\mathcal{C}_{\P^e}$, that is, $r^e$ is a rule of the form 
\[\langle c\leftarrow_i @^e[p_1,\dots,p_m;p_{m+1},\dots,p_n];\ \top\rangle\]
Its corresponding rule in the reduct $\P^e_{M^e}$, denoted as $r^e_{M^e}$, is then given by
\[\langle c\leftarrow_i \mathcal{B}_{M^e};\ \top\rangle\] being $\mathcal{B}_{M^e}=@^e[p_1,\dots,p_m;M^e(p_{m+1}),\dots,M^e(p_n)]$.

Notice that, $N^e$ satisfies the rule $r^e_{M^e}$ if and only if $\top\preceq N^e(c)\Dotted\leftarrow_i N^e(\mathcal{B}_{M^e})=c\leftarrow_i N^e(\mathcal{B}_{M^e})$. As a result, we can assert that $N^e$ satisfies the rule $r^e_{M^e}$ if and only if $N^e(\mathcal{B}_{M^e})\preceq c$.

Concerning the rule $r^{\tilde{e}}\in\P^{\tilde{e}}$, by Definition~\ref{def:correspEMALP}, it is given by
\[\langle p_\bot\leftarrow_i f_\bot(\neg p_\bot)\adjoint f_c(@^e[p_1,\dots,p_m;p_{m+1},\dots,p_n]); \top\rangle\]
Consequently, we obtain that the corresponding rule of $r^{\tilde{e}}$ in the reduct $\P^{\tilde{e}}_{M^{\tilde{e}}}$, denoted as $r^{\tilde{e}}_{M^{\tilde{e}}}$, is given by
\[\langle p_\bot\leftarrow_i f_\bot(M^{\tilde{e}}(\neg p_\bot))\adjoint f_c(\mathcal{B}_{M^{\tilde{e}}}); \top\rangle\]
being $\mathcal{B}_{M^{\tilde{e}}}=@^e[p_1,\dots,p_m;M^{\tilde{e}}(p_{m+1}),\dots,M^{\tilde{e}}(p_n)]$. Since  $p_i\neq p_\bot$, for each $i\in\{m+1,\dots,n\}$, we deduce that $M^e(p_i)=M^{\tilde{e}}(p_i)$, and thus $\mathcal{B}_{M^e}=\mathcal{B}_{M^{\tilde{e}}}$. Moreover, by definition of the interpretation $M^{\tilde{e}}$, $M^{\tilde{e}}(p_\bot)=\bot$, and therefore the following chain of equalities holds 
\[f_\bot(\hat{M}^{\tilde{e}}(\neg p_\bot))=f_\bot(\Dotted\neg M^{\tilde{e}}( p_\bot))=f_\bot(\Dotted\neg \bot)=f_\bot(\top)=\top\]
Hence, since $\adjoint$ satisfies the boundary condition with the top element,   the rule $r^{\tilde{e}}_{M^{\tilde{e}}}$ is semantically equivalent to
\[\langle p_\bot\leftarrow_i f_c(\mathcal{B}_{M^e}); \top\rangle\]

Therefore, the  interpretation $N^{\tilde{e}}$ satisfies   the rule $r^{\tilde{e}}_{M^{\tilde{e}}}$ if and only if $\top\preceq N^{\tilde{e}}(p_\bot)\leftarrow_i f_c(N^{\tilde{e}}(\mathcal{B}_{M^e}))$. Due to $p_\bot$ does not appear in $\mathcal{B}_{M^e}$, we deduce that $N^e(\mathcal{B}_{M^e})=N^{\tilde{e}}(\mathcal{B}_{M^e})$, whence it is followed that $N^{\tilde{e}}$ satisfies   the rule $r^{\tilde{e}}_{M^{\tilde{e}}}$ if and only if $f_c(N^e(\mathcal{B}_{M^e}))\preceq N^{\tilde{e}}(p_\bot)=\bot$. That is, $f_c(N^e(\mathcal{B}_{M^e}))=\bot$. Hence, according to the definition of the mapping $f_c$, we can assert that $N^{\tilde{e}}$ satisfies   the rule $r^{\tilde{e}}_{M^{\tilde{e}}}$ if and only if $N^e(\mathcal{B}_{M^e})\preceq c$, or equivalently, $N^e$ satisfies the rule $r^e_{M^e}$.

Consequently, we conclude that an interpretation $N^e$ is a model of the program $\P^e_{M^e}$ if and only if $N^{\tilde{e}}$ is a model of the program $\P^{\tilde{e}}_{M^{\tilde{e}}}$.  In particular, $M^e$ is a model of  $\P^e_{M^e}$ if and only if $M^{\tilde{e}}$ is a model of $\P^{\tilde{e}}_{M^{\tilde{e}}}$.

Now, we will demonstrate that, given an interpretation $M$, we have that $M^e$ is the least model of $\P^e_{M^e}$ if and only if $M^{\tilde{e}}$ is the least model of $\P^{\tilde{e}}_{M^{\tilde{e}}}$, which will be proved by reductio ad absurdum.

Suppose that $M^{\tilde{e}}$ is the least model of $\P^{\tilde{e}}_{M^{\tilde{e}}}$ but $M^e$ is not the least model of $\P^e_{M^e}$. Then, there exists $N^e\colon\Pi_{\P^e}\to L$ with $N^e\sqsubset M^e$ such that $N^e$ is a model of $\P^e_{M^e}$. Therefore, by the property proved above, the interpretation $N^{\tilde{e}}\colon\Pi_{\P^e}\cup\{p_\bot\}\to L$ given by $N^{\tilde{e}}(p)=N^e(p)$ if $p\in\Pi_{\P^e}$ and $N^{\tilde{e}}(p_\bot)=\bot$ is a model of $\P^{\tilde{e}}_{M^{\tilde{e}}}$, and the inequality $N^{\tilde{e}}\sqsubset M^{\tilde{e}}$ is straightforwardly satisfied. Therefore, we obtain a contradiction, since $M^{\tilde{e}}$ is the least model of $\P^{\tilde{e}}_{M^{\tilde{e}}}$ by hypothesis.

Finally, suppose that $M^e$ is the least model of $\P^e_{M^e}$ but $M^{\tilde{e}}$ is not the least model of $\P^{\tilde{e}}_{M^{\tilde{e}}}$, that is, there exists a model $N^{\tilde{e}}\colon \Pi_{\P^e}\cup\{p_\bot\}\to L$ of $\P^{\tilde{e}}_{M^{\tilde{e}}}$ verifying $N^{\tilde{e}}\sqsubset M^{\tilde{e}}$. 
Since, $N^e$ is equal to the restriction of $N^{\tilde{e}}$ to $\Pi_{\P^e}$, that is, 
$N^e=N^{\tilde{e}}_{|\Pi_{\P^e}}$ and $M^{\tilde{e}}(p_{\bot})=\bot=N^{\tilde{e}}(p_{\bot})$, we obtain that 
$$
N^e=N^{\tilde{e}}_{|\Pi_{\P^e}} \sqsubset M^{\tilde{e}}_{|\Pi_{\P^e}} =M^e
$$
 
Hence, according to the fact that $N^{\tilde{e}}$ is a model of $\P^{\tilde{e}}_{M^{\tilde{e}}}$, we obtain that $N^e$ is a model of $\P^e_{M^e}$, and therefore $M^e$ is not the least model of $\P^e_{M^e}$, in contradiction with the hypothesis.

Hence, we conclude that $M^e$ is the least model of $\P^e_{M^e}$ if and only if $M^{\tilde{e}}$ is the least model of $\P^{\tilde{e}}_{M^{\tilde{e}}}$, that is, $M^e$ is a stable model of $\P^e$ if and only if $M^{\tilde{e}}$ is a stable model of $\P^{\tilde{e}}$.
\qed
\end{proof}

Given an EMALP $\P^e$, one can wonder if there exist stable models of the constraint-free EMALP $\P^{\tilde{e}}$ such that the propositional symbol $p_\bot$ is not assigned to the bottom element in the lattice $L$. In that case, those stable models are not included in the characterization provided in Theorem~\ref{th:EMALP}.

Nevertheless, from Definition~\ref{def:correspEMALP}, we can deduce that any stable model maps the element $p_\bot$ to the bottom element in the lattice $L$, as the following result shows.

\begin{theorem}\label{prop:EMALP}
Let $\P^e$ be an EMALP and $\P^{\tilde{e}}$ be the corresponding constraint-free EMALP of $\P^e$. If $M^{\tilde{e}}\colon \Pi_{\P^e}\cup\{p_\bot\}\to L$ is a stable model of $\P^{\tilde{e}}$, then $M^{\tilde{e}}(p_\bot)=\bot$.
\end{theorem}
\begin{proof}
Suppose that $M^{\tilde{e}}\colon \Pi_{\P^e}\cup\{p_\bot\}\to L$ is a stable model of $\P^{\tilde{e}}$. First and foremost, notice that, if $\mathcal{C}_{\P^e}=\varnothing$, then there are no rule in $\P^{\tilde{e}}$ with head $p_\bot$, and thus neither are in the reduct $\P^{\tilde{e}}_{M^{\tilde{e}}}$. Therefore, as $M^{\tilde{e}}$ is the least model of $\P^{\tilde{e}}_{M^{\tilde{e}}}$, we straightforwardly obtain that $M^{\tilde{e}}(p_\bot)=\bot$.

Now, we assume that $\mathcal{C}_{\P^e}\neq\varnothing$. Clearly, as far as $\hat{M}^{\tilde{e}}(\neg p_\bot)$ is concerned, only the next two options are feasible: $\hat{M}^{\tilde{e}}(\neg p_\bot)=\bot$ or $\bot\prec\hat{M}^{\tilde{e}}(\neg p_\bot)$. From each of them, we will deduce that $M^{\tilde{e}}(p_\bot)=\bot$.

Suppose that $\hat{M}^{\tilde{e}}(\neg p_\bot)=\bot$. Then, according to Equation~\eqref{eq:corr_reduct}, for each rule $r^{\tilde{e}}$ in $\P^{\tilde{e}}$ of the form\footnote{Since $\mathcal{C}_{\P^e}\neq\varnothing$, we can ensure the existence of at least one rule of the form of $r^{\tilde{e}}$.}
\[\langle p_\bot\leftarrow_i f_\bot(\neg p_\bot)\adjoint f_c(@^e[ p_1, \dots, p_m; p_{m+1},\dots, p_n]); \top\rangle\]
its corresponding rule $r^{\tilde{e}}_{M^{\tilde{e}}}$ in the reduct $\P^{\tilde{e}}_{M^{\tilde{e}}}$ is defined as
\[\langle p_\bot\leftarrow_i f_\bot(\bot)\adjoint f_c(@^e[p_1,\dots,p_m;M^{\tilde{e}}(p_{m+1}),\dots,M^{\tilde{e}}(p_n)]); \top\rangle\]
Equivalently, as $f_\bot(\bot)=\bot$ and $\adjoint$ is a conjunctor of the extended multi-adjoint lattice, we obtain that $\bot\Dotted\adjoint x=\bot$ for each $x\in L$. Hence, the rule $r^{\tilde{e}}_{M^{\tilde{e}}}$ in $\P^{\tilde{e}}_{M^{\tilde{e}}}$   can  be rewritten as
\[r^{\tilde{e}}_{M^{\tilde{e}}}:\quad\langle p_\bot\leftarrow_i \bot; \top\rangle\]
Finally, as $p_\bot$ only appears in the rules in $\P^{\tilde{e}}_{M^{\tilde{e}}}$ which come from constraints and $M^{\tilde{e}}$ is the least model of $\P^{\tilde{e}}_{M^{\tilde{e}}}$, we conclude that\footnote{Notice that in this case, the lattice $L$ should be a singleton.}
\begin{eqnarray*}
M^{\tilde{e}}(p_\bot)&=&\inf\{c\in L\mid \top\preceq c\Dotted{\leftarrow_i} M^{\tilde{e}}(\bot)\}=\inf\{c\in L\mid \top\preceq c\Dotted{\leftarrow_i} \bot\}\\
&=&\inf\{c\in L\mid \top\Dotted{\adjoint_i}\bot\preceq c\}=\inf\{c\in L\mid \bot\preceq c\}=\inf L=\bot
\end{eqnarray*}

Now, suppose that $\bot\prec\hat{M}^{\tilde{e}}(\neg p_\bot)$. By definition of $f_\bot$, we obtain $f_\bot(\hat{M}^{\tilde{e}}(\neg p_\bot))=\top$. As a result, according to Equation~\eqref{eq:corr_reduct} we can assert that given a rule $r^{\tilde{e}}$ in $\P^{\tilde{e}}$ of the form
\[\langle p_\bot\leftarrow_i f_\bot(\neg p_\bot)\adjoint f_c(@^e[ p_1, \dots, p_m; p_{m+1},\dots, p_n]); \top\rangle\]
its corresponding rule $r^{\tilde{e}}_{M^{\tilde{e}}}$ in the reduct $\P^{\tilde{e}}_{M^{\tilde{e}}}$ is defined as
\[\langle p_\bot\leftarrow_i f_c(@^e[p_1,\dots,p_m;M^{\tilde{e}}(p_{m+1}),\dots,M^{\tilde{e}}(p_n)]); \top\rangle\]
Suppose now that the body of the rule $r^{\tilde{e}}_{M^{\tilde{e}}}$ is equal to $\top$, that is
\[f_c(@^e[p_1,\dots,p_m;M^{\tilde{e}}(p_{m+1}),\dots,M^{\tilde{e}}(p_n)])=\top\]
Then the rule $r^{\tilde{e}}_{M^{\tilde{e}}}$ can be rewritten as
\[\langle p_\bot\leftarrow_i \top; \top\rangle\]
As $M^{\tilde{e}}$ is the least model of $\P^{\tilde{e}}_{M^{\tilde{e}}}$, it satisfies the rule $r^{\tilde{e}}_{M^{\tilde{e}}}$, and thus
\[\top\preceq M^{\tilde{e}}(p_\bot)\Dotted\leftarrow_i \top\]
Making the corresponding computations and considering the adjoint conjunctor $\adjoint_i$ of the implication, we deduce that
\[\top=\top\Dotted{\adjoint_i} \top\preceq M^{\tilde{e}}(p_\bot)\]
This leads us to a contradiction, since, according to the fact that $\neg$ is a negator operator, the chain $\bot\prec \hat{M}^{\tilde{e}}(\neg p_\bot)=\Dotted\neg M^{\tilde{e}}(p_\bot)$ implies that $M^{\tilde{e}}(p_\bot)\neq\top$. Consequently, we can ensure that
\[f_c(@^e[p_1,\dots,p_m;M^{\tilde{e}}(p_{m+1}),\dots,M^{\tilde{e}}(p_n)])=\bot\]
for each rule $r^{\tilde{e}}_{M^{\tilde{e}}}$ in $\P^{\tilde{e}}_{M^{\tilde{e}}}$ of the form
\[\langle p_\bot\leftarrow_i f_c(@^e[p_1,\dots,p_m;M^{\tilde{e}}(p_{m+1}),\dots,M^{\tilde{e}}(p_n)]); \top\rangle\]
Hence, we obtain that the rule $r^{\tilde{e}}_{M^{\tilde{e}}}$ is equivalent  to
\[\langle p_\bot\leftarrow_i \bot; \top\rangle\]
Therefore, following an analogous reasoning to the previous one, we can assert that any stable model $M^{\tilde{e}}$ of the program $\P^{\tilde{e}}$ verifies $M^{\tilde{e}}(p_\bot)=\bot$.
\qed
\end{proof}

Theorems~\ref{th:EMALP} and~\ref{prop:EMALP} highlights the close connection between an EMALP and its corresponding constraint-free EMALP given in Definition~\ref{def:correspEMALP}. Indeed, one can ensure that the stable models of an EMALP are equivalent to the stable models of its corresponding constraint-free EMALP.

A straightforward outcome from these previous results is the fact that the number of stable models of $\P^e$ is equal to the number of stable models of $\P^{\tilde{e}}$. This fact gives rise to the next result, which concerns the existence and uniqueness of stable models.

\begin{corollary}\label{cor:existenceEMALP}
Let $\P^e$ be an EMALP and $\P^{\tilde{e}}$ the corresponding constraint-free EMALP of $\P^e$. Then, the following statements hold:
\begin{itemize}
\item There exists a stable model of $\P^e$ if and only if there exists a stable model of $\P^{\tilde{e}}$.
\item There exists a unique stable model of $\P^e$ if and only if there exists a unique stable model of $\P^{\tilde{e}}$.
\end{itemize}
\end{corollary}

In the next example, we carry out the translation detailed in Definition~\ref{def:correspEMALP} in order to obtain a constraint-free EMALP from the EMALP given in Example~\ref{ex:EMALP}.

\begin{example}\label{ex:EMALPwC}
Coming back to Example~\ref{ex:EMALP} and  considering the same EMALP $\P^e$, we have  
by the procedure in Definition~\ref{def:correspEMALP}, that  the rules $r_1^e$, $r_2^e$, $r_4^e$ and $r_5^e$ are included in the corresponding constraint-free EMALP $\P^{\tilde{e}}$ of $\P^e$. In what regards the rules $r_3^e$, it is  transformed  in the rule:
\[\langle p_\bot\leftarrow_\text{\L} f_\bot(\neg_1 p_\bot)\adjoint_G f_{0.7}(@^e_3[p;q,s,t])\ ;1\rangle\]
Notice that, we have arbitrarily chosen the conjunction $\adjoint_G$, but we can make use of any different conjunction whenever it satisfies the boundary condition with $\top$.

The constraint-free EMALP $\P^{\tilde{e}}$ is then given by the following four rules and one fact
\begin{equation*}
\begin{array}{ll}
r_1^{\tilde{e}}:\ \langle p\leftarrow_\text{P}@^e_1[p,q;s,t] \ \ ;\ 0.6\rangle &
r_4^{\tilde{e}}:\ \langle s\leftarrow_\text{G} @^e_4[p,q,s,t]\ ;\ 0.8\rangle\\
r_2^{\tilde{e}}:\ \langle q\leftarrow_\text{P} @^e_2[p,q;s,t]\ ;\ 0.5\rangle &
r_5^{\tilde{e}}:\ \langle t\leftarrow_\text{G} @^e_5[p,q,s,t]\ ;\ 0.8\rangle\\
r_3^{\tilde{e}}:\ \langle p_\bot\leftarrow_\text{\L} f_\bot(\neg_1 p_\bot)\adjoint_G f_{0.7}(@^e_3[p;q,s,t])\ ;1\rangle &
\end{array}
\end{equation*}
Now, according to Theorem~\ref{th:EMALP}, due to $N^e\equiv\{(p,\nicefrac{9}{85}),(q,0.36),(s,0.8),(t,0.8)\}$ is a stable model of $\P^e$, we can assert that the interpretation 
\[N^{\tilde{e}}\equiv\{(p,\nicefrac{9}{85}),(q,0.36),(s,0.8),(t,0.8),(p_\bot,0)\}\]
is a stable model of $\P^{\tilde{e}}$.
\qed
\end{example}

To summarize, given an EMALP $\P^e$, Definition~\ref{def:correspEMALP} provides a method in order to obtain a constraint-free EMALP $\P^{\tilde{e}}$ whose stable models coincide with the stable models of $\P^e$. One of the most interesting consequences of this method is that, if one knows that there exists at least a stable model (resp. a unique stable model) of $\P^{\tilde{e}}$, then the existence of stable models (resp. a unique stable model) of $\P^e$ is guaranteed.
However,  there are no results related to the existence or the uniqueness of stable models for either EMALPs or constraint-free EMALP. 

Recently, these kind of results were introduced in the particular case of normal  residuated logic programming in~\cite{Madrid:2012} and in the general case of the  multi-adjoint framework
 in~\cite{fss:manlp2017}. Therefore, we should try to introduce a transformation mechanism from constraint-free EMALPs to   MANLPs, which preserve the semantics (the stable models).  As a consequence, we may apply the already introduced results in  constraint-free EMALPs, and by the transformation already given in this section they can also be considered in  the general case of EMALPs.

\section{Translating constraint-free EMALP into MANLP}\label{sec:final}

This section will present a transformation from  a constraint-free EMALP to  a MANLP,  such that 
 the stable models of the original constraint-free EMALP are the same as the stable models of the obtained MANLP. The underlying idea of this translation method is provided next.

Let $\P^{\tilde{e}}$ be a constraint-free EMALP and $\mathcal{N}_{\P^{\tilde{e}}}$ be the set of propositional symbols that appear in an order-reversing argument of an extended aggregator in the body of some rule in $\P^{\tilde{e}}$. Given an operation symbol $\neg$ associated with an involutive negation operator $\Dotted\neg\colon L\to L$, suppose that for each $q\in\mathcal{N}_{\P^{\tilde{e}}}$ there exists a propositional symbol $\text{not}_q$ such that any stable model $M^{\tilde{e}}$ of $\P^{\tilde{e}}$ verifies the equality $M^{\tilde{e}}(\text{not}_q)=\hat{M}^{\tilde{e}}(\neg q)$. As $\Dotted\neg$ is an involutive negation operator, we obtain that $\Dotted\neg M^{\tilde{e}}(\text{not}_q)=\Dotted\neg \hat{M}^{\tilde{e}}(\neg q)=\Dotted\neg\Dotted\neg M^{\tilde{e}}(q)=M^{\tilde{e}}(q)$.

Therefore, for each rule $\langle p\leftarrow_i @^e[ p_1, \dots, p_m; p_{m+1},\dots, p_n]; \vartheta \rangle\in\P^{\tilde{e}}$, we deduce that
\[@^e[ p_1, \dots, p_m; M^{\tilde{e}}(p_{m+1}),\dots, M^{\tilde{e}}(p_n)]=@^e[ p_1, \dots, p_m; \Dotted\neg M^{\tilde{e}}(\text{not}_{p_{m+1}}),\dots,\Dotted \neg M^{\tilde{e}}(\text{not}_{p_n})]\]
This fact leads us to assert that the rule
\[\langle p\leftarrow_i @^e[ p_1, \dots, p_m; p_{m+1},\dots, p_n]; \vartheta \rangle\]
can be rewritten as the rule
\[\langle p\leftarrow_i @[ p_1, \dots, p_m, \text{not}_{p_{m+1}},\dots, \text{not}_{p_n}]; \vartheta \rangle\]
where the mapping $\Dotted @\colon L^n\to L$ is defined, for all $\vartheta_1,\dots, \vartheta_n\in L$,  as
\[\Dotted{@}[\vartheta_1, \dots, \vartheta_m, \vartheta_{m+1},\dots, {\vartheta_n}]=\Dotted{@^e}[\vartheta_1,\dots,\vartheta_m;\Dotted\neg {\vartheta_{m+1}},\dots,\Dotted\neg \vartheta{_{n}}]
\]

Notice that,   as $\Dotted {@^e}$ is an extended aggregator, it is an order-preserving mapping in the first $m$ arguments, and so $\Dotted @$ is order-preserving in the first $m$ arguments. Furthermore, since $\Dotted {@^e}$ is an order-reversing mapping in the last $n-m$ arguments and $\Dotted \neg$ is also an order-reversing mapping, we obtain that $\Dotted @$ is an order-preserving mapping in the last $n-m$ arguments. Hence, we conclude that $\Dotted @$ is an aggregator. 
Therefore, 
the rule
\[\langle p\leftarrow_i @[ p_1, \dots, p_m, \text{not}_{p_{m+1}},\dots, \text{not}_{p_n}]; \vartheta \rangle\]
is a   weighted rule without negations  and so, a particular case of rule allowed in a MANLP.

Last but not least, suppose that the assumption of existing, for each $q\in\mathcal{N}_{\P^{\tilde{e}}}$, a propositional symbol $\text{not}_q$ such that any stable model $M^{\tilde{e}}$ of $\P^{\tilde{e}}$ verifies the equality $M^{\tilde{e}}(\text{not}_q)=\hat{M}^{\tilde{e}}(\neg q)$ does not hold. In that case, we include the next rule in $\P$ for each propositional symbol $q\in\mathcal{N}_{\P^{\tilde{e}}}$:
\[r_q:\qquad\langle \text{not}_q\leftarrow_j \neg q; \top\rangle\]
It is straightforward that the rule $r_q$ can belong to a MANLP, for each $q\in\mathcal{N}_{\P^{\tilde{e}}}$. Furthermore, any stable model $M$ of $\P$ is, by definition, the least model of the reduct $\P_M$. Since $r_q$ is the unique rule in $\P$ with head $\text{not}_q$, the interpretation of $\text{not}_q$ under $M$ is equal to $\hat{M}(\neg q)$, as it will be demonstrated in Proposition~\ref{prop:EMALPwC}.

The following definition collects the previous comments in order to formally introduce the proposed transformation. 

\begin{definition}\label{def:correspEMALPwC}
Let $\P^{\tilde{e}}$ be a constraint-free EMALP and $\mathcal{N}_{\P^{\tilde{e}}}$ the set of propositional symbols that appear in an order-reversing argument of an extended aggregator in the body of some rule in $\P^{\tilde{e}}$. Given the symbol $\neg$ associated with  an involutive negation operator $\Dotted\neg\colon L\to L$, the corresponding MANLP $\P$ of $\P^{\tilde{e}}$ is defined as the following set of rules
\begin{eqnarray*}
\P&=&\big\{\langle p\leftarrow_i @[ p_1, \dots, p_m, \text{not}_{p_{m+1}},\dots, \text{not}_{p_n}]; \vartheta \rangle\mid \\
& &\langle p\leftarrow_i @^e[ p_1, \dots, p_m; p_{m+1},\dots, p_n]; \vartheta \rangle\in\P^{\tilde{e}}\big\}\\
& \cup & \big\{\langle \text{not}_q\leftarrow_j \neg q; \top\rangle\mid q\in\mathcal{N}_{\P^{\tilde{e}}}\big\}
\end{eqnarray*}
where $i,j\in\{1,\dots,n\}$, $\text{not}_q\notin\Pi_{\P^{\tilde{e}}}$, for each $q\in\mathcal{N}_{\P^{\tilde{e}}}$, and $\Dotted{@}\colon L^n\to L$ is an aggregator which is defined, for all $\vartheta_1,\dots, \vartheta_n\in L$,  as
\[\Dotted{@}[\vartheta_1, \dots, \vartheta_m, \vartheta_{m+1},\dots, {\vartheta_n}]=\Dotted{@^e}[\vartheta_1,\dots,\vartheta_m;\Dotted\neg {\vartheta_{m+1}},\dots,\Dotted\neg \vartheta{_{n}}]
\]
\end{definition}

Given an interpretation $M\colon L^n\to L$, the reduct of $\P$ with respect to $M$ is defined analogously to the procedure explained in Section~\ref{sec:MANLP} (Equation~\eqref{eq:reductMANLP}).

Now, the stable models of the MANLP $\P$ will be proved to be, as expected, equivalent to the stable models of the EMALP $\P^{\tilde{e}}$, improving and complementing the  results given in~\cite{Janssen2012}. To reach this conclusion, two results will be introduced. The first one provides a characterization of the stable models of $\P^{\tilde{e}}$ in terms of a family of stable models of $\P$.

\begin{theorem}\label{th:EMALPwC}
Let $\P^{\tilde{e}}$ be a constraint-free EMALP and $\P$ the corresponding MANLP of $\P^{\tilde{e}}$,  $M^{\tilde{e}}\colon\Pi_{\P^{\tilde{e}}}\to L$ an interpretation, and  $M\colon \Pi_{\P^{\tilde{e}}}\cup\{\text{not}_q\mid q\in\mathcal{N}_{\P^{\tilde{e}}}\}\to L$ given by $M(p)=M^{\tilde{e}}(p)$ if $p\in\Pi_{\P^{\tilde{e}}}$ and $M(\text{not}_q)=\hat{M}(\neg q)$ for all $q\in\mathcal{N}_{\P^{\tilde{e}}}$.  Then, $M^{\tilde{e}}$ is a stable model of $\P^{\tilde{e}}$ if and only if   $M$ is a stable model of $\P$.
\end{theorem}
\begin{proof}
Given two interpretations $M^{\tilde{e}}\colon\Pi_{\P^{\tilde{e}}}\to L$ and $N^{\tilde{e}}\colon\Pi_{\P^{\tilde{e}}}\to L$, we define $N_M\colon \Pi_{\P^{\tilde{e}}}\cup\{\text{not}_q\mid q\in\mathcal{N}_{\P^{\tilde{e}}}\}\to L$ as $N_M(p)=N^{\tilde{e}}(p)$ if $p\in\Pi_{\P^{\tilde{e}}}$ and $N_M(\text{not}_q)=\hat{M}^{\tilde{e}}(\neg q)$ for each $q\in\mathcal{N}_{\P^{\tilde{e}}}$. 
In the particular case of $M^{\tilde{e}}=N^{\tilde{e}}$, we will have that $M_M\colon \Pi_{\P^{\tilde{e}}}\cup\{\text{not}_q\mid q\in\mathcal{N}_{\P^{\tilde{e}}}\}\to L$ is defined as $M_M(p)=M^{\tilde{e}}(p)$ if $p\in\Pi_{\P^{\tilde{e}}}$ and $M_M(\text{not}_q)=\hat{M}^{\tilde{e}}(\neg q)=\Dotted \neg\hat{M}^{\tilde{e}}(q)=\Dotted \neg {M}_M(q)=  \hat{M_M} (\neg q)$ for each $q\in\mathcal{N}_{\P^{\tilde{e}}}$, and we simply write $M$ instead of $M_M$.

First of all, we will prove  that $N^{\tilde{e}}$ is a model of the reduct $\P^{\tilde{e}}_{M^{\tilde{e}}}$ if and only if $N_M$ is a model of the reduct $\P_{M}$.
The first step for getting this equivalence  will be to prove  that an interpretation $N\colon \Pi_{\P^{\tilde{e}}}\cup\{\text{not}_q\mid q\in\mathcal{N}_{\P^{\tilde{e}}}\}\to L$ satisfies the rule
\[
\langle \text{not}_q\leftarrow_j \hat{M}(\neg q); \top\rangle
\]
in the reduct $\P_{M}$ if and only if $M(\text{not}_q)\preceq N(\text{not}_q)$.
Since the following chain straightforwardly holds
\begin{eqnarray*}
 \top\Dotted{\adjoint_j} N(\hat{M}(\neg q))=N(\hat{M}(\neg q))=\hat{M}(\neg q)=
 M(\text{not}_q)
\end{eqnarray*}
and   $(\adjoint_j,\leftarrow_j)$ is an adjoint pair, we obtain that
\[
\top= \hat{N}(\text{not}_q\leftarrow_j\hat{M}(\neg q)) \quad \hbox{iff}\quad
M(\text{not}_q)= \top\Dotted{\adjoint_j} N(\hat{M}(\neg q)) \preceq  N(\text{not}_q)
\]
which leads to the proof of this first claim.

Consequently, in particular, as $N_M(\text{not}_q)=\hat{M}^{\tilde{e}}(\neg q)=M(\text{not}_q)$, the interpretation $N_M$ satisfies the rules with  head $ \text{not}_q$ in the reduct $\P_{M}$.
Now, notice that a rule 
\[\langle p\leftarrow_i @^e[ p_1, \dots, p_m; p_{m+1},\dots, p_n]; \vartheta \rangle\]
  belongs to the program $\P^{\tilde{e}}$ if and only if the rule
\[\langle p\leftarrow_i @[ p_1, \dots, p_m, \text{not}_{p_{m+1}},\dots, \text{not}_{p_n}]; \vartheta \rangle\]
belongs to $\P$, and thus, due to $@$ is order-preserving in every argument, this is equivalent to  this rule belongs to the reduct $\P_{M}$. Therefore, regarding the reducts $\P^{\tilde{e}}_{M^{\tilde{e}}}$ and $\P_{M}$, we deduce that the rule
\[\langle p\leftarrow_i @^e[ p_1, \dots, p_m; M^{\tilde{e}}(p_{m+1}),\dots, M^{\tilde{e}}(p_n)]; \vartheta \rangle\]
belongs to $\P^{\tilde{e}}_{M^{\tilde{e}}}$ if and only if
\[
\langle p\leftarrow_i @[ p_1, \dots, p_m, \text{not}_{p_{m+1}},\dots, \text{not}_{p_n}]; \vartheta \rangle
\]
belongs to $\P_{M}$. Taking into account this fact, we will see that $N^{\tilde{e}}$ is a model of $\P^{\tilde{e}}_{M^{\tilde{e}}}$ if and only if $N_M$ is a model of the reduct $\P_{M}$. Indeed, the interpretation $N_M$ satisfies the rule 
\[
\langle p\leftarrow_i @[ p_1, \dots, p_m, \text{not}_{p_{m+1}},\dots, \text{not}_{p_n}]; \vartheta \rangle
\]
if and only if
\[\vartheta\preceq \hat{N_M}(p\leftarrow_i @[ p_1, \dots, p_m, \text{not}_{p_{m+1}},\dots, \text{not}_{p_n}]\big)
\]
that is
\[
\vartheta\preceq N_M(p)\Dotted\leftarrow_i \Dotted{@}[ N_M(p_1), \dots, N_M(p_m), N_M(\text{not}_{p_{m+1}}),\dots, N_M(\text{not}_{p_n})]\]
which is equivalent, according to the definition of $@$, to the inequality
\[
\vartheta\preceq N_M(p)\Dotted\leftarrow_i \Dotted{@^e}[ N_M(p_1), \dots, N_M(p_m);\Dotted\neg N_M(\text{not}_{p_{m+1}}),\dots,\Dotted\neg N_M(\text{not}_{p_n})]
\]
Furthermore, as $N_M(p)=N^{\tilde{e}}(p)$ and $N_M(\text{not}_q)=\Dotted\neg M^{\tilde{e}}(q)$, for each $q\in\mathcal{N}_{\P^{\tilde{e}}}$, we obtain that $N_M$ satisfies the inequality above if and only if $N^{\tilde{e}}$ satisfies the following one
\[
\vartheta\preceq N^{\tilde{e}}(p)\Dotted\leftarrow_i \Dotted {@^e}[ N^{\tilde{e}}(p_1), \dots, N^{\tilde{e}}(p_m);\Dotted \neg \Dotted\neg M^{\tilde{e}}(p_{m+1}),\dots, \Dotted\neg \Dotted\neg M^{\tilde{e}}(p_n)]\big)
\]
Taking into account that $\Dotted\neg$ is by hypothesis an involutive negation, the last inequality is equivalent to
\begin{equation}\label{eq:th_EMALPwC}
\vartheta\preceq N^{\tilde{e}}(p)\Dotted\leftarrow_i \Dotted {@^e}[ N^{\tilde{e}}(p_1), \dots, N^{\tilde{e}}(p_m); M^{\tilde{e}}(p_{m+1}),\dots, M^{\tilde{e}}(p_n)]\big)
\end{equation}
Furthermore, as $N^{\tilde{e}}(x)=x$ for each $x\in L$ and $M^{\tilde{e}}(p_i)\in L$ for each $i\in\{m+1,\dots,n\}$, we can assert that $N^{\tilde{e}}(M^{\tilde{e}}(p_i))=M^{\tilde{e}}(p_i)$, for each $i\in\{m+1,\dots,n\}$. Therefore, Equation~\eqref{eq:th_EMALPwC} can be rewritten as
\[
\vartheta\preceq N^{\tilde{e}}(p)\Dotted\leftarrow_i \Dotted {@^e}[ N^{\tilde{e}}(p_1), \dots, N^{\tilde{e}}(p_m); N^{\tilde{e}}(M^{\tilde{e}}(p_{m+1})),\dots, N^{\tilde{e}}(M^{\tilde{e}}(p_n))]\big)
\]
Equivalently
\[\vartheta\preceq \hat{N}^{\tilde{e}}(p\leftarrow_i @^e[ p_1, \dots, p_m; M^{\tilde{e}}(p_{m+1}),\dots,M^{\tilde{e}}(p_n)]\big)\]
Hence, we conclude that $N_M$ satisfies the rule 
\[
\langle p\leftarrow_i @[ p_1, \dots, p_m, \text{not}_{p_{m+1}},\dots, \text{not}_{p_n}]; \vartheta \rangle
\]
in the reduct $\P_{M}$ if and only if $N^{\tilde{e}}$ satisfies the rule
\[
\langle p\leftarrow_i @^e[ p_1, \dots, p_m; M^{\tilde{e}}(p_{m+1}),\dots, M^{\tilde{e}}(p_n)]; \vartheta \rangle
\]
in the program $\P^{\tilde{e}}_{M^{\tilde{e}}}$. 
Therefore, $N^{\tilde{e}}$ is a model of $\P^{\tilde{e}}_{M^{\tilde{e}}}$ if and only if $N_M$ is a model of the reduct $\P_{M}$. 

To finish with this demonstration, we will see that $M^{\tilde{e}}$ is  {the least} model of $\P^{\tilde{e}}_{M^{\tilde{e}}}$ if and only if $M$ is  {the least} model of $\P_{M}$ by reductio ad absurdum. In fact, suppose that $M$ is the least model of $\P_{M}$ but there exists $N^{\tilde{e}}\colon\Pi_{\P^{\tilde{e}}}\to L$ with $N^{\tilde{e}}\sqsubset M^{\tilde{e}}$ such that $N^{\tilde{e}}$ is a model of $\P^{\tilde{e}}_{M^{\tilde{e}}}$. Then, by the previous proved equivalence,  the interpretation $N_M\colon \Pi_{\P^{\tilde{e}}}\cup\{\text{not}_q\mid q\in\mathcal{N}_{\P^{\tilde{e}}}\}\to L$, given by $N_M(p)=N^{\tilde{e}}(p)$ if $p\in\Pi_{\P^{\tilde{e}}}$ and $N_M(\text{not}_q)=\hat{M}^{\tilde{e}}(\neg q)$ otherwise, is a model of $\P_{M}$. As a consequence, we obtain that $N_M\sqsubset M$ is straightforwardly satisfied and so, we obtain that $M$ is not the least model of $\P_{M}$, in contradiction with the hypothesis.

Finally, suppose that $M^{\tilde{e}}$ is the least model of $\P^{\tilde{e}}_{M^{\tilde{e}}}$ but $M$ is not the least model of $\P_{M}$. As a result, we can assert that there exists a model $N\colon \Pi_{\P^{\tilde{e}}}\cup\{\text{not}_q\mid q\in\mathcal{N}_{\P^{\tilde{e}}}\}\to L$ of $\P_{M}$ verifying $N\sqsubset M$. Then, there exists $p\in \Pi_{\P^{\tilde{e}}}\cup\{\text{not}_q\mid q\in\mathcal{N}_{\P^{\tilde{e}}}\}$ such that $N(p)\prec M(p)$. Notice that, if $p\in \{\text{not}_q\mid q\in\mathcal{N}_{\P^{\tilde{e}}}\}$, that is, there exists $q\in\mathcal{N}_{\P^{\tilde{e}}}$ such that $p=\text{not}_q$, then the inequality $N(\text{not}_q)\prec M(\text{not}_q)=\hat{M}^{\tilde{e}}(\neg q)$ holds, which contradicts that $N$ satisfies the rule $\langle \text{not}_q\leftarrow_j \hat{M}(\neg q); \top\rangle$
in the reduct $\P_{M}$, as we showed at the beginning of this proof.  As a consequence, we also have that $N(\text{not}_q)=\hat{M}^{\tilde{e}}(\neg q)$, for all $q\in\mathcal{N}_{\P^{\tilde{e}}}$.

Hence, we can assert that there exists $p'\in \Pi_{\P^{\tilde{e}}}$ such that $N(p')\prec M(p')$. 
As a result, the interpretation $N^{\tilde{e}}\colon\Pi_{\P^{\tilde{e}}}\to L$ given by $N^{\tilde{e}}(p)=N(p)$, for all $p\in \Pi_{\P^{\tilde{e}}}$, satisfies $N^{\tilde{e}}\sqsubset M^{\tilde{e}}$. Since $N$ is a model of $\P_{M}$, with $N(\text{not}_q)=\hat{M}^{\tilde{e}}(\neg q)$, for all $q\in\mathcal{N}_{\P^{\tilde{e}}}$, we deduce that $N^{\tilde{e}}$ is a model of $\P^{\tilde{e}}_{M^{\tilde{e}}}$, and thus $M^{\tilde{e}}$ is not the least model of $\P^{\tilde{e}}_{M^{\tilde{e}}}$, in contradiction with the hypothesis.

Therefore, we conclude that $M^{\tilde{e}}$ is the least model of $\P^{\tilde{e}}_{M^{\tilde{e}}}$ if and only if $M$ is the least model of $\P_{M}$, that is, $M^{\tilde{e}}$ is a stable model of $\P^{\tilde{e}}$ if and only if $M$ is a stable model of $\P$.
\qed
\end{proof}

Theorem~\ref{th:EMALPwC} establishes that each stable model of $\P^{\tilde{e}}$ is equivalent to a stable model of $\P$ in the set 
\[
S=\{M\in\mathcal{I}_{\mathfrak{L}}\mid M(\text{not}_q)=\hat{M}(\neg q),\hbox{ for all }q\in\mathcal{N}_{\P^{\tilde{e}}}\}
\]

In the following, we will see that any stable model of the MANLP $\P$ belongs to $S$. As a result, any stable model of $\P$ is taken into account in Theorem~\ref{th:EMALPwC}, and thus we can conclude that the set of stable models of $\P^{\tilde{e}}$ actually coincide with the set of stable models of $\P$.

\begin{theorem}\label{prop:EMALPwC}
Let $\P^{\tilde{e}}$ be a constraint-free EMALP, $\P$ be its corresponding MANLP and $\mathcal{N}_{\P^{\tilde{e}}}$ be the set of propositional symbols that appear in an order-reversing argument position of an extended aggregator in the body of some rule in $\P^{\tilde{e}}$. Then, any stable model $M$ of the MANLP $\P$ satisfies $M(\text{not}_q)=\hat{M}(\neg q)$, for all $q\in\mathcal{N}_{\P^{\tilde{e}}}$.
\end{theorem}
\begin{proof}
Given  a stable model $M$ of the MANLP $\P$ and  $q\in\mathcal{N}_{\P^{\tilde{e}}}$, in particular, $M$ satisfies the rule
 \[\langle \text{not}_q\leftarrow_j \hat{M}(\neg q); \top\rangle\]
 in the reduct $\P_{M}$.  Hence, from the first equivalence proved in the proof of Theorem~\ref{th:EMALPwC}, we deduce that  $\hat{M}(\neg q)\preceq M(\text{not}_q)$.  On the other hand, 
 according to the fact that $M$ is the least model of $\P_{M}$ and $\langle \text{not}_q\leftarrow_j \hat{M}(\neg q); \top\rangle$ is the unique rule with head $\text{not}_q$ in $\P_{M}$, we have that $M(\text{not}_q)$ takes the least value such that $\hat{M}(\neg q)\preceq M(\text{not}_q)$. Hence, we can assert that $M(\text{not}_q)= \hat{M}(\neg q)$, and this equality holds for each $q\in\mathcal{N}_{\P^{\tilde{e}}}$.
\qed
\end{proof}

From  Theorems~\ref{th:EMALPwC} and~\ref{prop:EMALPwC}, we can assert that the stable models of an EMALP coincide with the stable models of its transformed MANLP. As a consequence, the number of stable models of $\P^{\tilde{e}}$ is equal to the number of stable models of $\P$.

This fact leads us to deduce that there exists at least a stable model of $\P^{\tilde{e}}$ if and only if there exists at least a stable model of $\P$. An equivalent outcome is obtained regarding the uniqueness of stable models of  $\P^{\tilde{e}}$ and $\P$. These results are formalized as follows.

\begin{corollary}\label{cor:existenceEMALPwC}
Let $\P^{\tilde{e}}$ be a constraint-free EMALP and $\P$ the corresponding MANLP of $\P^{\tilde{e}}$. Then, the following statements hold:
\begin{itemize}
\item There exists a stable model of $\P^{\tilde{e}}$ if and only if there exists a stable model of $\P$.
\item There exists a unique stable model of $\P^{\tilde{e}}$ if and only if there exists a unique stable model of $\P$.
\end{itemize}
\end{corollary}

As a continuation of Examples~\ref{ex:EMALP} and~\ref{ex:EMALPwC}, we will complete the translation from an EMALP into a MANLP with the same stable models.

\begin{example}
Consider the EMALP $\P^e$ defined in Example~\ref{ex:EMALP}
and its corresponding constraint-free EMALP $\P^{\tilde{e}}$ given in Example~\ref{ex:EMALPwC}.
Notice that, the propositional symbols that appear in an order-reversing argument of an extended aggregator in the body of the rules in $\P^{\tilde{e}}$ are $s,t$ (rules $r_1^{\tilde{e}}$ and $r_2^{\tilde{e}}$) and $q,p_\bot$ (rule $r_3^{\tilde{e}}$). Hence, we obtain that $\mathcal{N}_{\P^{\tilde{e}}}=\{q,s,t,p_\bot\}$.

According to Definition~\ref{def:correspEMALPwC}, in order to define the corresponding MANLP $\P$ of $\P^{\tilde{e}}$ we only need an involutive negation in $[0,1]$. For the sake of simplicity, as the negation $\neg_1$ is an involutive negation, we will make use of this operator to define the MANLP $\P$. Hence, we obtain that the corresponding MANLP $\P$ of $\P^{\tilde{e}}$ is defined as the following seven rules and one fact
\begin{equation*}
\begin{array}{ll}
r_1:\ \langle p\leftarrow_\text{P}@_1[p,q,\text{not}_{s},\text{not}_{t}] \ \ ;\ 0.5\rangle &
r_6:\ \langle \text{not}_{q}\leftarrow_\text{G} \neg_1 q\ ;\ \top\rangle\\
r_2:\ \langle q\leftarrow_\text{P} @_2[p,q,\text{not}_{s},\text{not}_{t}]\ ;\ 0.6\rangle &
r_7:\ \langle \text{not}_{s}\leftarrow_\text{G} \neg_1 s\ ;\ \top\rangle\\
{r_3:\ \langle p_\bot\leftarrow_\text{\L} @_3[p,\text{not}_{q},\text{not}_{p_\bot}]\ ;1\rangle} &
r_8:\ \langle \text{not}_{t}\leftarrow_\text{G} \neg_1 t\ ;\ \top\rangle\\
r_4:\ \langle s\leftarrow_\text{G} @_4[p,q,s,t]\ ;\ 0.8\rangle &
r_9:\ \langle \text{not}_{p_\bot}\leftarrow_\text{G} \neg_1 p_\bot\ ;\ \top\rangle\\
r_5:\ \langle t\leftarrow_\text{G} @_5[p,q,s,t]\ ;\ 0.8\rangle &
\end{array}
\end{equation*}
where the aggregator operators $@_1,@_2,@_4,@_5\colon[0,1]^4\to[0,1]$ and $@_3\colon[0,1]^5\to[0,1]$ are defined as 
\begin{equation*}
\begin{array}{lllll}
@_1[p,q,\text{not}_{s},\text{not}_{t}]&=&@^e_1[p,q;\neg_1(\text{not}_{s}),\neg_1(\text{not}_{t})]&=&\min\big\{\frac{q}{\neg_1(\text{not}_{s})+\neg_1(\text{not}_{t})+0.1},1\big\}\\
@_2[p,q,\text{not}_{s},\text{not}_{t}]&=&@^e_2[p,q;\neg_1(\text{not}_{s}),\neg_1(\text{not}_{t})]&=&\max\big\{\neg_1(\neg_1(\text{not}_{s}))), \neg_2(\neg_1(\text{not}_{t}))\big\}
\\
\end{array}
\end{equation*}
\begin{equation*}
\begin{array}{rcl}
@_3[p,\text{not}_{q},\text{not}_{p_\bot}]&=&f_\bot(\neg_1(\neg_1(\text{not}_{p_\bot})))\adjoint_G f_{0.7}(\neg_1(\neg_1(\text{not}_{q})))
\\
@_4[p,q,s,t]&=&@^e_4[p,q,s,t]=1\\
@_5[p,q,s,t]&=&@^e_5[p,q,s,t]=\max\{s,0.7\}
\end{array}
\end{equation*}

Since $N^{\tilde{e}}\equiv\{(p,\nicefrac{9}{85}),(q,0.36),(s,0.8),(t,0.8),(p_\bot,0)\}$
is a stable model of the constraint-free EMALP $\P^{\tilde{e}}$, Theorem~\ref{th:EMALPwC} leads us to conclude that the interpretation
$N\equiv\{(p,\nicefrac{9}{85}),(q,0.36),(s,0.8),(t,0.8),(p_\bot,0),(\text{not}_{q},0.64),$ $(\text{not}_{s},0.2),(\text{not}_{t},0.2),(\text{not}_{p_\bot},1)\}$
is a stable model of the MANLP $\P$.
\qed
\end{example}

Due to the fact that the semantics of an extended multi-adjoint logic program is defined in terms of the stable models of the program, ensuring the existence of stable models becomes a crucial task in order to define its semantics.

According to Corollary~\ref{cor:existenceEMALPwC}, we obtain that any
result related to the existence (resp. unicity) of stable models for MANLPs
is likely to be used in order to guarantee the existence (resp. unicity) of
stable models of the original EMALP. Indeed, Theorem~\ref{thm:existencia} leads us to infer the following result.

\begin{theorem}\label{th:existEMALP}
Let $(K,\preceq,\leftarrow_1,\adjoint_1,\dots,\leftarrow_n,\adjoint_n, @_1^e, \dots, @_k^e)$ be an extended multi-adjoint lattice where $K$ is a non-empty convex compact set in an euclidean space and $\P^{\tilde{e}}$ a finite constraint-free EMALP defined on this lattice. If $\Dotted\adjoint_1,\dots, \Dotted\adjoint_n$ and the extended aggregator operators in the body of the rules of  $\P^{\tilde{e}}$ are continuous operators, and there exists a continuous involutive negation $\Dotted\neg\colon K\to K$ then $\P^{\tilde{e}}$  has at least a stable model.
\end{theorem}
\begin{proof}
Let $\P$ be the corresponding MANLP of $\P^{\tilde{e}}$ given in Definition~\ref{def:correspEMALPwC} by means of the negation operator $\neg$. The MANLP $\P$ is then defined on the multi-adjoint normal lattice $(K,\preceq,\leftarrow_1,\adjoint_1,\dots,\leftarrow_n,\adjoint_n,\neg)$, being $\Dotted\adjoint_1,\dots, \Dotted\adjoint_n$ and $\Dotted\neg$ continuous operators. Clearly, since $\Dotted\neg$ is by hypothesis a continuous mapping, then the aggregator operator in the body of the rule
\[\langle \text{not}_q\leftarrow_j \neg q; \top\rangle\]
is a continuous operator, for each $q\in\mathcal{N}_{\P^{\tilde{e}}}$.

Furthermore, for each rule 
\[\langle p\leftarrow_i @^e[ p_1, \dots, p_m; p_{m+1},\dots, p_n]; \vartheta \rangle\]
in $\P^{\tilde{e}}$, taking into account that $\Dotted{@^e}$ and $\Dotted\neg$ are continuous mappings, we obtain that the aggregator $\Dotted{@}\colon L^n\to L$ defined as
\[\Dotted{@}[\vartheta_1, \dots, \vartheta_m,  {\vartheta_{m+1}},\dots, {\vartheta_n}]=\Dotted {@^e}[\vartheta_1,\dots,\vartheta_m;\Dotted\neg  {\vartheta_{m+1}},\dots,\Dotted\neg  {\vartheta_{n}}]
\]
for all $\vartheta_1,\dots, \vartheta_n\in L$, 
is a continuous mapping. Therefore, the aggregator operator $\Dotted{@}$  associated with the  symbol $@$  in the body of the rule
\[\langle p\leftarrow_i @[ p_1, \dots, p_m, \text{not}_{p_{m+1}},\dots, \text{not}_{p_n}]; \vartheta \rangle\]
is a continuous mapping.

Hence, we can apply Theorem~\ref{thm:existencia} to the MANLP $\P$, from which we obtain that $\P$ has at least a stable model $M$. Then, given a stable model $M$ of $\P$, Proposition~\ref{prop:EMALPwC} allows us to assert that $M(\text{not}_q)=\hat{M}(\neg q)$, for each $q\in\mathcal{N}_{\P^{\tilde{e}}}$. Finally, as stated by Theorem~\ref{th:EMALPwC}, we conclude that the interpretation $M^{\tilde{e}}\colon\Pi_{\P^{\tilde{e}}}\to L$ given by $M^{\tilde{e}}(p)=M(p)$, for each $p\in\Pi_{\P^{\tilde{e}}}$, is a stable model of $\P^{\tilde{e}}$. Thus, there exists at least a stable model of $\P^{\tilde{e}}$.
\qed
\end{proof}

In the following, an illustrative example shows how Theorem~\ref{th:existEMALP} can be used to ensure the existence of stable models for a general EMALP.

\begin{example}
Suppose that we remove the rule $r_3^e$ in the EMALP $\P^e$ given in Example~\ref{ex:EMALP}, constructing a new EMALP ${\P^e}^*$. This may be interesting if, for instance, some features of the problem that is being simulated by the EMALP change and these changes imply that the condition of the models of $\P$ satisfying that the evaluation of $q$ is greater than $0.3$ is not demanded anymore. The EMALP ${\P^e}^*$ is then defined on the multi-adjoint lattice $([0,1],\leq,\leftarrow_\text{G},\adjoint_G,\leftarrow_\text{P},\adjoint_\text{P},\leftarrow_\text{\L},\adjoint_\text{\L})$ as the following four rules:
\begin{equation*}
\begin{array}{ll}
r_1^e:\ \langle p\leftarrow_\text{P}@^e_1[p,q;s,t] \ ;\ 0.5\rangle\quad &
r_4^e:\ \langle s\leftarrow_\text{G} @^e_4[p,q,s,t]\ ;\ 0.8\rangle\\
r_2^e:\ \langle q\leftarrow_\text{P} @^e_2[p,q;s,t]\ ;\ 0.6\rangle &
r_5^e:\ \langle t\leftarrow_\text{G} @^e_5[p,q,s,t]\ ;\ 0.8\rangle\\
\end{array}
\end{equation*}
being $@^e_1,@^e_2,@^e_4,@^e_5\colon[0,1]^4\to[0,1]$  defined in Example~\ref{ex:EMALP}.

Observe that $[0,1]$ is a non-empty convex compact set in the euclidean space $([0,1],+,*,\mathbb{R})$, being $+$ and $*$ the usual sum and product in $\mathbb{R}$, respectively. Furthermore, the conjunctions $\adjoint_G$ and $\adjoint_\text{P}$ and the extended aggregators $@^e_1$, $@^e_2$, $@^e_4$ and $@^e_5$ are continuous operators in $[0,1]^4$. Taking into account that, for instance, the mapping $\neg\colon[0,1]\to[0,1]$ given by $\neg(x)=1-x$ is a continuous involutive negation, Theorem~\ref{th:existEMALP} leads us to conclude that there exists at least a stable model of ${\P^e}^*$. For instance, the interpretation $N^e$ given by $N^e\equiv\{(p,\nicefrac{9}{85}),(q,0.36), (s,0.8), (t,0.8)\}$ is a stable model of ${\P^e}^*$. 
\qed
\end{example}

Therefore, the introduced results in this section have completed the transformation from a general EMALP to a semantically equivalent MANLP. Specifically, each stable model of the transformed  MANLP provides another stable model of the original EMALP.

\section{Conclusions and future work}\label{sec:conclusion}

An extension of multi-adjoint normal logic programming has been presented. In this new kind of programs, special rules called constraints have been included and aggregator operators with order-reversing arguments are allowed to appear in the body of the rules. 
This consideration allows, for instance, to consider 
multi-adjoint normal logic programs with multiple negations. 

Moreover, this extension generalizes the one given in~\cite{Janssen2012,Madrid2017} and   considerably increases the flexibility of MANLPs.
After presenting the syntax and the semantics of EMALPs, a procedure in order to translate an EMALP into a constraint-free EMALP, preserving the semantic given by the stable models, has been provided. Then, a method to simulate a constraint-free EMALP by means of a semantically equivalent MANLP has been presented. These two procedures make possible, for example, that a user can consider results associated with stable models of MANLPs in order to obtain information about the stable models of an EMALP. For instance, we have shown that  Theorem~\ref{thm:existencia}, presented in~\cite{fss:manlp2017}, provides
  sufficient conditions under which the existence of stable models of a constraint-free EMALP can be ensured.  The uniqueness results given in~\cite{fss:manlp2017} can also be applied. 
   
Since the auxiliary mappings $f_c$ considered in the transformation from EMALPs to constraint-free EMALPs  are non-continuous, for each
$c \in L$, Theorem 6 cannot be used in order to guarantee the existence
of stable models for an EMALP with constraints.  Therefore, new results need to be studied in the future.
For example, Madrid proposes in~\cite{Madrid2017} a feasible procedure to come to this aim for residuated logic programs on the unit interval with constraints which can be  generalized to the case of EMALPs. This extension is one of the proposals for further research. 
Moreover, we have other two feasible alternatives to be considered as a future work:
\begin{itemize}
\item Obtaining a transformation from EMALPs to EMALPs without constraints by means of continuous operations, which would allow us to make use of Theorem~\ref{thm:existencia}.
\item Providing sufficient conditions that ensure the existence of stable models of MANLPs and do not require the continuity of the operators in the body of the rules.
\end{itemize}

Finally, the study of inconsistency and incoherent information~\cite{buchman2017,iwann2017,eusflatMANLP2017,NMadrid3,madrid:2011,merhej2017,redl2017}    in EMALPs will be another important task in the future.

\end{document}